\documentclass[aps,reprint,pra,superscriptaddress,showpacs]{revtex4-1}

\usepackage{graphicx}
\usepackage{amssymb}
\usepackage{amsmath}
\usepackage{amsfonts}
\usepackage{appendix}

\begin{document}

\preprint{APS/123-QED}

\title{Quantum information scrambling in non-Markovian open quantum systems}

\author{Li-Ping Han}
\affiliation{Key Laboratory of Advanced Optoelectronic Quantum Architecture and Measurement, Ministry of Education,
School of Physics, Beijing Institute of Technology, Beijing 100081, China}
\affiliation{School of Science, Tianjin University of Technology, Tianjin 300384, China}
\author{Jian Zou}
\email{zoujian@bit.edu.cn}
\affiliation{Key Laboratory of Advanced Optoelectronic Quantum Architecture and Measurement, Ministry of Education,
School of Physics, Beijing Institute of Technology, Beijing 100081, China}
\author{Hai Li}
\affiliation{School of Information and Electronic Engineering, Shandong Technology and Business University,
Yantai 264005, China}
\author{Bin Shao}
\affiliation{Key Laboratory of Advanced Optoelectronic Quantum Architecture and Measurement, Ministry of Education,
School of Physics, Beijing Institute of Technology, Beijing 100081, China}

\begin{abstract}
In this paper we investigate the dynamics of a spin chain whose two end spins interact with two independent non-Markovian baths by using the non-Markovian quantum state diffusion (QSD) equation approach.  Specifically two issues about quantum information scrambling in open quantum system are addressed. The first issue is that tripartite mutual information (TMI) can quantify information scrambling properly via its negative value in closed system, whether it is still suitable to indicate  quantum scrambling in open quantum system. However we find that negative TMI is not an suitable quantifier of information scrambling in open quantum system in some cases while negative tripartite logarithmic negativity (TLN) is more appropriate. The second one is that up to now almost all the open quantum system effects on information scrambling reported were focus on the Markovian environment, while the effect of non-Markovian environment on information scrambling is still elusive. Significantly our results show that the memory effect of environment will be beneficial to the emergence of quantum information scrambling. Moreover, it is found that  environment is generally detrimental  for information scrambling in a long time, while in some cases it will be helpful for information scrambling in a short time.

%
\end{abstract}

\maketitle


\section{\label{sec:I}INTRODUCTION}
Entanglement, as a key resource in quantum information processing, is believed to give significant insights into physical mechanisms in a variety of fields \cite{1,2,3,4,5}. How quantum information stored in local degrees of freedom of  initial state of a many-body system propagates and distributes over the global degrees of freedom of the  system, which is known as information scrambling, is an interesting topics from the fundamental point of view \cite{6}. And it stimulates a broad range of research interest in various fields, for example, quantum information \cite{7,8}, high energy physics \cite{9,10}, quantum-thermodynamics \cite{11,12} and condensed matter physics \cite{13,14}, etc. Scrambling is generically identified as  delocalization of quantum information \cite{15,16,17,18} in a many-body system. A general accepted measure of information scrambling is the so-called out-of-time-order correlator (OTOC), which is associated with the growth of the square commutator between two initially commuting observables \cite{15,19,20,21,22,23,24}.

Besides OTOC, tripartite mutual information (TMI) can also be a probe of information scrambling \cite{7}, which becomes negative if quantum
  information is delocalized, i.e., information is shared globally rather than in a bipartite manner.  A particular advantage for TMI is that the information measure does not rely on any selection of operators. The only choice is the partitioning of the Hilbert space which decides whether information is distributed or not \cite{25}. It has been proven that TMI is essentially equivalent to OTOC in capturing the feature of information scrambling by means of the channel-state duality \cite{7}, while they capture different aspects of quantum dynamics \cite{26}. Recently, more attention has been paid to  TMI in many-body quantum systems.  At first, temporal TMI has been investigated by using the channel-state duality in Refs. \cite{27,28}. Later, instantaneous TMI also has been used to study information scrambling in Refs. \cite{29,30,31}. The method  used in this paper is instantaneous TMI. When TMI is non-negative at some time, the information at this moment is localized, while at some time TMI is negative, the information is delocalized now. If TMI is non-negative at the beginning and becomes negative as time evolves, which means that information turns into delocalized, namely, scrambling occurs. The definition of TMI is based on the von Neumann entropy, whose important caveat is that it captures both quantum and classical correlations. It is thus important to isolate the quantum contribution to the entropy. To this end, tripartite logarithmic negativity (TLN) \cite{32} is analogously proposed to characterize the pure quantum information that is scrambled.

It is well known that realistic quantum systems inevitably interact with their surrounding environments, resulted in decoherence and dissipation. The time evolution of such an open quantum system is usually characterized by a quantum master equation through Markovian approximation  \cite{33}, corresponding to a memoryless environment, and this leads to a monotonic information flow from the system of interest to the environment. When the environment's memory can not be ignored, a backflow of information from the environment to the system occurs and the non-Markovian description of the system dynamics is required  \cite{34}. It has been found that non-Markovianity can lead to a significant variety of physical effects in the dynamics of open quantum systems \cite{35,36,37,38,39,40,41}, and can serve as a resource in information theory \cite{42,43,44,45,46}.  However, it is normally a hard task to solve the non-Markovian dynamics of the system and many theoretical approaches have been developed (see, e.g., Refs.  \cite{47,48,49,50,51,52,53,54,55,56,57,58,59,60,61,62}). Among these approaches, the non-Markovian quantum state diffusion (QSD) equation method  \cite{48,49,50,51,52,53} has been proven to be effective.

 It is noteworthy that quantum information scrambling is rooted in the spread of entanglement, which is hard to preserve in the presence of environment. The influence of environment noise on delocalization of information could not be neglected. Several works about open quantum system dynamics by using different quantifiers of information scrambling have been reported \cite{63,64,65,66,67,68,69,70,71,72}, such as corrected OTOC \cite{64,65}, a ratio of OTOC \cite{66}, mutual information \cite{67}, fidelity \cite{68}, etc. In Ref. \cite{65} it has been found that taking an open bipartite OTOC as a probe one can differentiate the contribution of scrambling from decoherence and also distinguish integrable dynamics from chaotic dynamics. And it has been found that dissipation and decoherence always suppress information scrambling for Markovian environment \cite{64,65}. It has been shown in Ref. \cite{66} that one can distinguish scrambling from decoherence in strongly interacting quantum systems by utilizing a teleportation-based decoding protocol. While Touil and Deffner have found that OTOC is not a good quantifier of information scrambling for open quantum systems, and they have related the competing effects of scrambling and decoherence to their respective contributions to the entropy change \cite{67}. Up to now, most of the works about information scrambling  by using TMI in the literature have focused on closed systems \cite{3,25,26,29,30,73} while TMI for open quantum system has not been fully considered. To our knowledge, there are so far three studies about information scrambling of open quantum systems by using TMI. In Refs. \cite{31} and \cite{74} Y. Li \textit{{et al}}. have proposed a collision model to simulate the information dynamics in an all-optical system and found that  non-Markovianity plays dual roles in affecting the dynamics of information. In Ref. \cite{75} Sur and  Subrahmanyam have  found that local quantum dynamical process can cause information scrambling even when the unitary evolution dynamics is non-scrambling in nature.  Similar to OTOC, whether TMI is a suitable quantifier of information scrambling for open quantum system or not is still an open question. Up to now, most of the open quantum system effects on information scrambling reported  have focused on the Markovian environment, the effect of non-Markovianity on quantum information scrambling is still elusive which requires further study.

 To address these questions, in this paper we focus on information delocalization in the presence of environment by using instantaneous TMI and TLN. The model we considered is a spin chain whose two end spins interact with two independent non-Markovian baths. We obtain the system's dynamics by using the QSD equation approach. Interestingly, we find that in some cases though TMI is negative entanglement might be zero, and thus negative TMI is not an appropriate probe of information scrambling in open quantum system, but negative TLN is. By comparing the dynamics of TLN with TMI, we can distinguish quantum information scrambling from the total information delocalization in open quantum system. Our results show that in general, environment is detrimental to information scrambling in a long time, while in some cases environment will be helpful for  the emergence of information scrambling  in a short time. More importantly we find that non-Markovianity plays a beneficial role in  both the total information delocalization and quantum information scrambling.

\section{\label{sec:II}MODEL AND METHODS}
The system we consider in this paper is a one dimensional $XXZ$ spin chain which consists of $N$ qubits, and the Hamiltonian is
\begin{equation}
 H_s  = \sum\limits_{i = 1}^{N - 1} {J_{i,i + 1} \left( {\sigma _i^x \sigma _{i + 1}^x  + \sigma _i^y \sigma _{i + 1}^y+ \Delta \sigma _i^z \sigma _{i + 1}^z } \right)},\label{1}  \\
  \\
 \end{equation}
where $J_{i,i + 1}$ is the coupling strength between the nearest neighbor sites $i$ and $i + 1$, and $\sigma _j \left( {j = x,y,z} \right)$ are the Pauli operators. Here we take  $J_{i,i + 1}  =  - 1$ throughout. For $\Delta  = 1$, it is the isotropic Heisenberg chain, which is the interacting case and  can be solved by the  Bethe ansatz \cite{76}. While for $\Delta  = 0$, it reduces to the $XX$ chain and  can be mapped to a free fermion model \cite{77}. This Hamiltonian Eq. (1) is integrable, and the dynamics of such integrable system can be understood by the propagation of quasi-particles, entangling different regions of the system as they propagate \cite{78}. Information that is initially localized in some region is spread by these quasi-particles, which move at different velocities. Thus, this information will disperse, leading in general to delocalized information among subsystems.

We suppose that the two end spins of the chain interact with two baths $H_{{\rm{1b}}}^{}$ and $H_{{\rm{2b}}}^{}$ respectively ( see Fig. 1). The total Hamiltonian  can be written as
\begin{equation}
H_{\rm{tot}}  = H_{\rm{s}}  + \sum\limits_{j = 1,2} {H_{jb} }  + H_{{\rm{int}}},\label{2}
\end{equation}
with the free Hamiltonian for the left and right bosonic bath $H_{jb}^{}\left({j = 1,2}\right)$
\begin{equation}
H_{jb}= \sum\limits_k {\omega _{jk} b_{jk}^{\dag} b_{jk} },\label{3}
\end{equation}
and the interaction described by
\begin{equation}
H_{\rm{int}}  = \sum\limits_{j = 1,2} {\sum\limits_k {\left( g_{jk}L_j  b_{jk}^\dag + g_{jk}^* L_j^\dag  b_{jk} \right)}}.\label{4}
\end{equation}
Here, $L_j$ is the Lindblad operator, $b_{jk}^\dag \left( b_{jk}\right)$
is the bosonic creation (annihilation) operator of the $k\rm{th}$ mode of the $j\rm{th}$ bath with frequency $\omega _{jk}$ and $g_{jk}$ is the coupling strength between the system and the $k\rm{th}$ mode of the $j\rm{th}$ bath.  We assume that the baths are both at zero temperature.

\begin{figure}[h!]
\centering
\includegraphics[width=8.3cm]{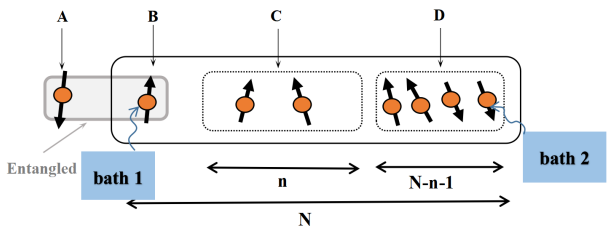}

\caption{A schematic of the model considered in this paper. The spin chain is divided into three parts B, C and D, where its two end spins interact with two baths, i.e., bath 1 and bath 2 respectively. Qubit A  is initially maximally entangled with qubit B, while C and D are not correlated with A and B initially. It is noticed that A is an ancillary qubit, which does not interact with the chain BCD.  }
\end{figure}

In order to investigate the effect of non-Markovianity on scrambling, we use the QSD equation approach \cite{49,53} and the evolution of the system can be derived in a form of time-local non-Markovian master equation (for the detail derivation see  Appendix A) \cite{79}
\begin{align}
\frac{\partial }{{\partial t}}&\rho _s  =  - i\left[ {H_s ,\rho _s } \right]\nonumber\\
&+ \sum\limits_{j = 1,2} {\left( {\left[ {L_j ,\rho _s \bar O_j^\dag  \left( t \right)} \right]} \right.}\left. { - \left[ {L_j^\dag  ,\bar O_j \left( t \right)\rho _s } \right]} \right), \label{5}
\end{align}
where $\bar O_j \left( {t} \right) = \int_0^t {\alpha _j \left( {t,s} \right)O_j } \left( {t,s } \right)ds$ and $O$ is an operator (for the detail definition see  Appendix A) \cite{80}.   $\alpha _j \left( {t,s} \right) = \frac{{\Gamma _j \gamma _j }}{2}e^{ - \gamma _j \left| {t - s} \right|}$ is the correlation function, which corresponds to the Ornstein-Uhlenbeck process \cite{47,48,81,82}. Here, $\Gamma _j$ denotes the coupling strength between the system and the $j\rm{th}$ bath. ${\raise0.7ex\hbox{$1$} \!\mathord{\left/{\vphantom {1 {\gamma _j }}}\right.\kern-\nulldelimiterspace}\!\lower0.7ex\hbox{${\gamma _j }$}}$ measures the correlation time between two separate time instances $t$ and $s$, which indicates the memory time of the $j\rm{th}$ bath.
When $\gamma _j \to \infty$, ${\raise0.7ex\hbox{$1$} \!\mathord{\left/{\vphantom {1 {\gamma _j }}}\right.\kern-\nulldelimiterspace}\!\lower0.7ex\hbox{${\gamma _j }$}} \to 0$, and in this case there is no correlation between the two separate time points $t$ and $s$, i.e., $\alpha _j \left( {t,s} \right) = \delta _j \left( {t,s} \right)$,  which represents the bath is  Markovian. Generally the Markov approximation is based on the assumption that the correlation time of the reservoir is much smaller than the time scale on which the system changes. When $\gamma _j$ is large enough, i.e., ${\raise0.7ex\hbox{$1$} \!\mathord{\left/{\vphantom {1 {\gamma _j }}}\right.\kern-\nulldelimiterspace}\!\lower0.7ex\hbox{${\gamma _j }$}}$ is small enough, the dynamics can be approximately regarded as Markovian. While when the parameter $\gamma _j$ is small, non-Markovian properties can be observed \cite{80,83,84,85,86}. As $\gamma _j$ is decreased a crossover from Markovian to non-Markovian dynamics can occur.  The operator $\bar O_j \left( t \right)$ satisfies  \cite{79}
\begin{align}
\frac{\partial }{{\partial t}}\bar O_1 \left( t \right) &= \frac{{\Gamma _1 \gamma _1 }}{2}L_1  - \gamma _1 \bar O_1 \left( t \right)\nonumber\\
&+ \left[ { - iH_s  - L_1^\dag  \bar O_1 \left( t \right)} \right.\left. { - L_2^\dag  \bar O_2 \left( t \right),\bar O_1 \left( t \right)} \right],\label{6}
\end{align}
\begin{align}
\frac{\partial }{{\partial t}}\bar O_2 \left( t \right) &= \frac{{\Gamma _2 \gamma _2 }}{2}L_2  - \gamma _2 \bar O_2 \left( t \right)\nonumber\\
 &+ \left[ { - iH_s  - L_1^\dag  } \right.\bar O_1 \left( t \right)\left. { - L_2^\dag  \bar O_2 \left( t \right),\bar O_2 \left( t \right)} \right].\label{7}
\end{align}
We can obtain the non-Markovian dynamics of the spin chain using Eqs. (5-7).

Next, we introduce the initial state used in this paper. Firstly, a product state between an ancillary qubit A and the system is prepared
\begin{align}
\frac{1}{{\sqrt 2 }}\left( {\left| 0 \right\rangle _A  + \left| 1 \right\rangle _A } \right) \otimes \left| \Xi  \right\rangle _{BCD}.\label{8}
\end{align}
Here, $\left| \Xi  \right\rangle _{BCD}$ is the initial state of system, which is divided into three parts B, C, and D (see Fig. 1). In this paper, we choose the initial system state as N\'{E}EL state $\left| \Xi  \right\rangle _{BCD}  = \left| {0101...01} \right\rangle$ and $\left| \Xi  \right\rangle _{BCD}  = \left| {00...00} \right\rangle$. Then a CNOT gate is applied on qubit A and qubit B, and in this way the information about A is locally encoded in B through the entanglement between A and B.

In this paper, we will consider the following two types of  Lindblad operators respectively. The first type corresponds to $L_j  = \sigma _j^z$ $(j = 1,2)$, which describes the dephasing process. For this Lindblad operator, the z-component of the total spins in the system is a conserved quantity $\left[ {\sum {\sigma _j^z ,H_s } } \right] = 0$. The second type is $L_j  = \sigma _j^ -$ $(j = 1,2)$, which describes the dissipative process, where $\sigma _j^ -$ denotes the lowering operator.

\section{\label{sec:III}RESULTS AND DISCUSSIONS}
\subsection{Effects of baths on information scrambling}

We next consider the effects of baths on information scrambling by instantaneous TMI.  TMI among the ancillary qubit A and the subsystems B, C is defined as
\begin{align}
I_3 \left( {A:B:C} \right) = I_2 \left( {A:B} \right) + I_2 \left( {A:C} \right) - I_2 \left( {A:BC} \right).\label{9}
\end{align}
$I_2 \left( {X:Y} \right) = S_X  + S_Y  - S_{XY}$ is bipartite mutual information (BMI) between $X$ and $Y$, which measures the total correlation (quantum and classical) between two subsystems of a composite system, and $S_X  =  - {\rm{Tr}}_X \left[ {\hat \rho _X } \right.\left. {\ln \hat \rho _X } \right]$ is the von Neumann entropy of the corresponding reduced density matrix $\hat \rho _X$.

From an information-theoretic point of view, TMI quantifies how the total (quantum and classical) information is shared among the subsystems A, B and C. $I_3 \left( {A:B:C} \right)$ is negative when $I_2 \left( {A:B} \right) + I_2 (A:C) < I_2 \left( {A:BC} \right)$, which implies that the sum of the total information that shared between A and B, A and C is smaller than that between A and BC together. In this case, the information about A is nonlocally stored in B and C such that measurements of B and C alone are not able to reconstruct A. Thus, a negative value of TMI is associated with  delocalization of the total information, or the total information being scrambled. If TMI is non-negative at the beginning and becomes negative with time evolution, it means that information turns into delocalized, namely, scrambling occurs.

 Firstly, we consider initial N\'{E}EL state and plot the time evolution of TMI in Figs. 2(b) and (c) for two different types of baths $L = \sigma ^ -$ and $L = \sigma ^z$ with $\Gamma  = 0.5$ respectively, while Fig. 2(a) is in the absence of baths ($\Gamma  = 0$) for comparison. It is shown in Fig. 2(a) that TMI can be negative, implying that the total information (quantum and classical) is scrambled inside BCD in the absence of baths, which is consistent with the result of Ref. [29]. Compared with Fig. 2(a), Fig. 2(b) shows that the maximum absolute value of the negative value of TMI for $L = \sigma ^ -$ becomes smaller and TMI gradually decays to zero in the presence of baths. It means that the total information is totally lost at last and delocalization of the total information only lasts for a finite time. In Fig. 2(c), the maximum absolute value of the negative value of TMI for $L = \sigma ^z$ becomes smaller and TMI decreases at first and finally arrives at a negative steady value. It is noted that the result is different from that of $L = \sigma ^ -$, more specifically for $L = \sigma ^z$ the information is not totally lost and there is residual information at last. From numerical calculation we find that the entanglement has disappeared when TMI reaches its steady value, which means that in this case there is no more quantum correlation, let alone quantum information scrambling. As is known, information scrambling is related to quantum correlation, and from above results we can learn that negative TMI does not always mean quantum information scrambling for open quantum system, because the residual information at last is purely classical in this case. An important caveat of the von Neumann entropy is that it captures both quantum and classical correlations. Then it is necessary to isolate the quantum contribution. To this end, we consider the bipartite logarithmic negativity (BLN), which is a proper measure of entanglement in mixed state, and its definition is \cite{87}
\begin{align}
\varepsilon _2  = \log \left( {\left| {\rho _{XY}^{T_Y } } \right|_1 } \right),\label{10}
\end{align}
where $\left| \Theta  \right|_1  = {\mathop{\rm Tr}\nolimits} \sqrt {\Theta ^\dag  \Theta }$ is the trace norm of $\Theta$ and $
\rho _{XY}^{T_Y }$ is the partial transpose of a density matrix. By replacing BMI on the right side of Eq. (9) with BLN, analogous to the quantity TMI, TLN is defined as \cite{32}

\begin{align}
\varepsilon _3 \left( {A:B:C} \right) = \varepsilon _2 \left( {A:B} \right) + \varepsilon _2 \left( {A:C} \right) - \varepsilon _2 \left( {A:BC} \right).\label{11}
\end{align}
A negative value of TLN implies delocalization of quantum information among A, B and C, while a non-negative value of TLN indicates quantum information mostly stored in bipartite partitions and  no delocalization  occurring.

In Fig. 3 we plot the time evolution of TLN for the same initial state.  Fig. 3(a) shows the time evolution of TLN in the absence of baths ($\Gamma  = 0$) for comparison, while, Figs. 3(b) and (c) are for $L = \sigma ^ -$ and $L = \sigma ^ z$ with $\Gamma  = 0.5$ respectively. The behavior of TLN shown in Fig. 3 is similar to that  in Fig. 2. TLN in Fig. 3(a) can be negative, which indicates that the quantum information is also scrambled in the absence of baths. Comparing Figs. 3(b) and (c) with Fig. 3(a),  we can see that the maximum absolute value of the negative value of TLN gets smaller and the duration of delocalization of quantum information is limited. Unlike TMI for $L = \sigma ^ z$ saturating to a negative steady value after a long time evolution, TLN (see Fig. 3(c)) decreases to zero at last, which means that finally quantum information is totally lost. Comparing Fig. 2 with Fig. 3, we can see that TMI lasts for a longer time than TLN in the presence of baths. Especially for $L = \sigma ^ z$, TMI saturates to a negative value after TLN decays to zero. It implies that when entanglement is zero, TMI can still be negative. Hence, negative TMI is not a good diagnosis of information scrambling for open quantum systems. By comparing the dynamics of TLN and TMI, we can distinguish quantum information scrambling from the total information delocalization in open quantum system. The decay of TLN to zero at last in the presence of baths shown in Fig. 3 implies that quantum information scrambling is suppressed by  these two different types of baths. Quantum information scrambling can only occur in a short time, and then disappear in a long time. This phenomenon can be understood as that the interaction between the system and bath creates entanglement between them, which in turn destroys the entanglement within the system, and hence diminishes delocalization of quantum information. It is noticed that though the environment has a negative impact on scrambling, there are regimes in which information is still distributed across all regions in Hilbert space in the early period. Comparing these two types of interactions, it can be seen that the time interval that TLN stays negative in the case of $L = \sigma ^ -$ is larger than that in the case of $L = \sigma ^ z$ for the same values of $\Gamma$ and $\gamma$. For $L = \sigma ^ z$  the total number of excitations for both ancillary qubit A and the system is conserved and thus the effective Hilbert subspace for quantum information is the same as that without baths. It is noticed that for $L = \sigma ^ -$ and initial N\'{E}EL state, at beginning due to partially decaying, the space belonging to each excitation  might be occupied which means that the effective Hilbert subspace is enlarged at the early time. While as time evolves further, the number of excitations gradually decreases and at last the system will evolve into the ground state completely, i.e., the size of effective Hilbert subspace after a transient period of time is gradually decreased. Although the total number of excitations is conserved in the case of $L = \sigma ^ z$,  decoherence occurs, which means that  the coherence  and quantum correlation gradually disappear as time evolves. And thus in which case $L = \sigma ^ z$ or $L = \sigma ^ -$  information scrambling  lasts for longer time depends on whether the coherence disappears faster or the  excitation  decays faster. For $XXZ$ chain, the coherence decays faster for $L = \sigma ^ z$ than  the excitation  decays for $L = \sigma ^ -$. On the other hand, it can be seen from Figs. 3(b) and (c) that the maximum absolute value of the negative value of TLN for $L = \sigma ^ -$ is larger than that for $L = \sigma ^ z$ because in this case as mentioned above the effective Hilbert subspace for $L = \sigma ^ -$ is larger than that for $L = \sigma ^ z$ in the transient period.

\begin{figure}[h!]
\centering
\includegraphics[width=8.3cm]{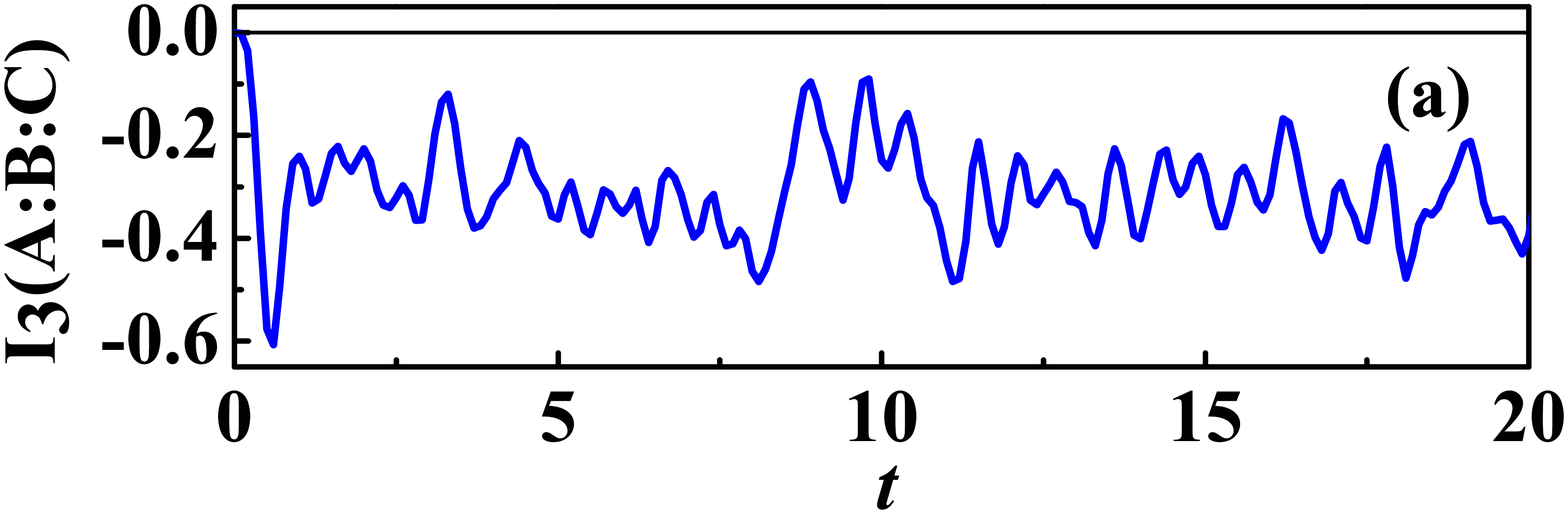}
\centering
\includegraphics[width=8.3cm]{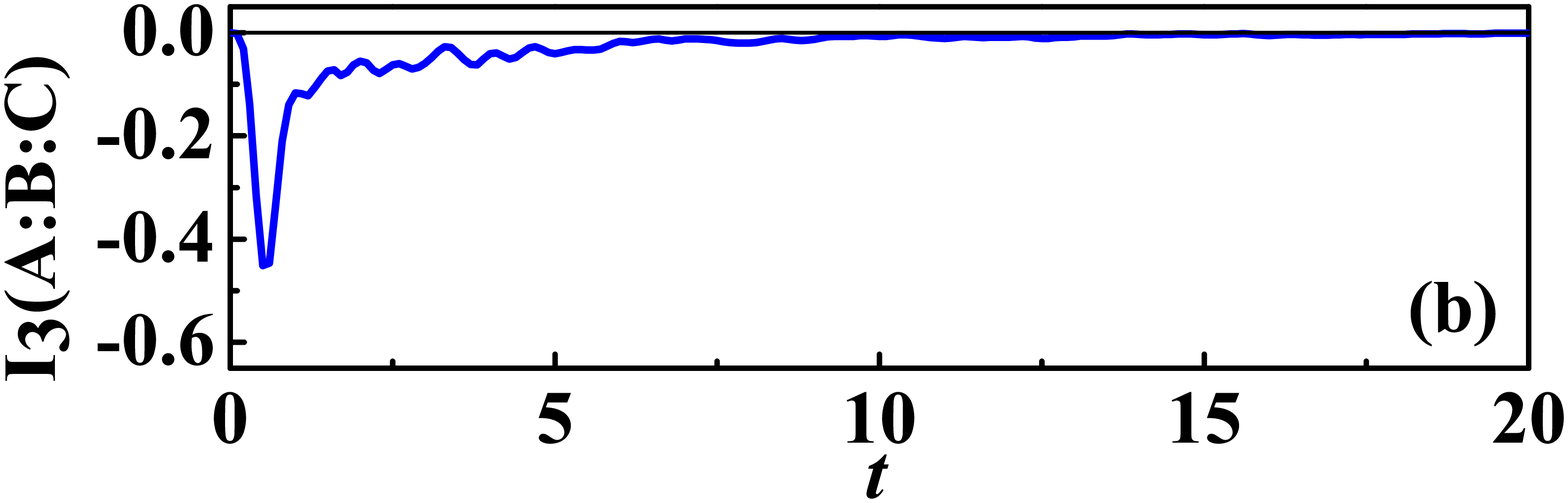}
\centering
\includegraphics[width=8.3cm]{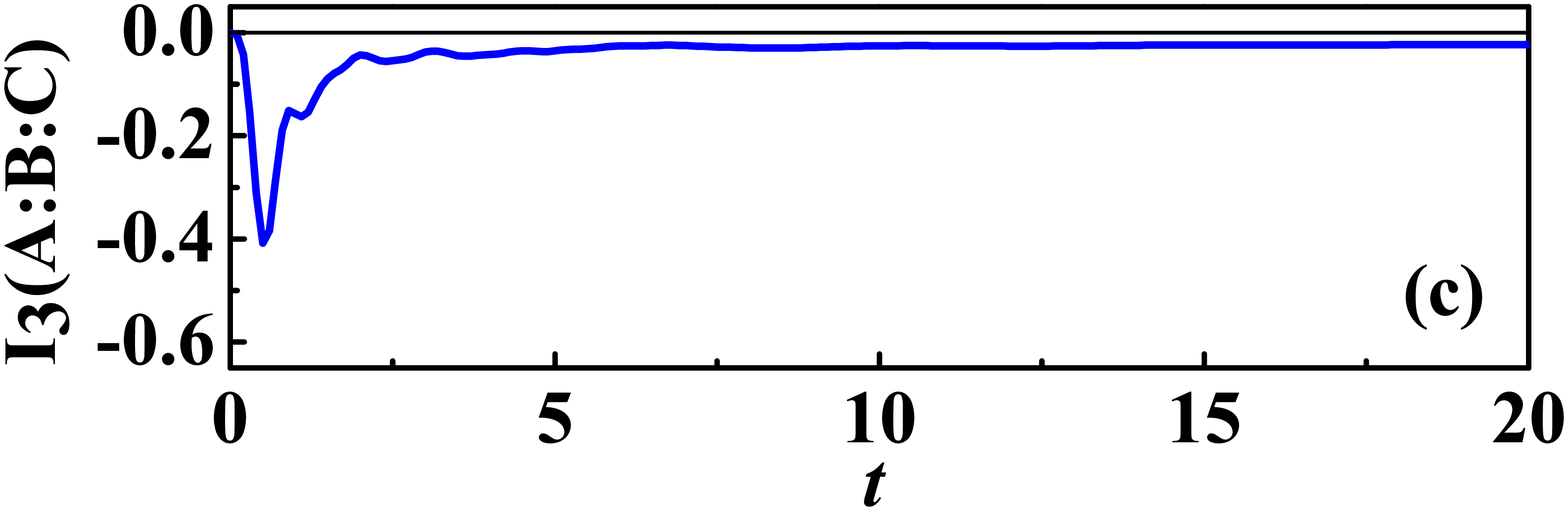}
\caption{TMI of $XXZ$ chain as a function of time for initial N\'{E}EL state, (a) in the absence of bath ($\Gamma  = 0$), (b) $L = \sigma ^ -$, and (c) $L = \sigma ^ z$. For both (b) and (c), $\Gamma _1  = \Gamma _2  = \Gamma {\rm{ = }}0.5$, and $\gamma _1  = \gamma _2  = \gamma {\rm{ = }}5$. Here $N = 6$, $n = 2$.}
\end{figure}

\begin{figure}[h!]
\centering
\includegraphics[width=8.3cm]{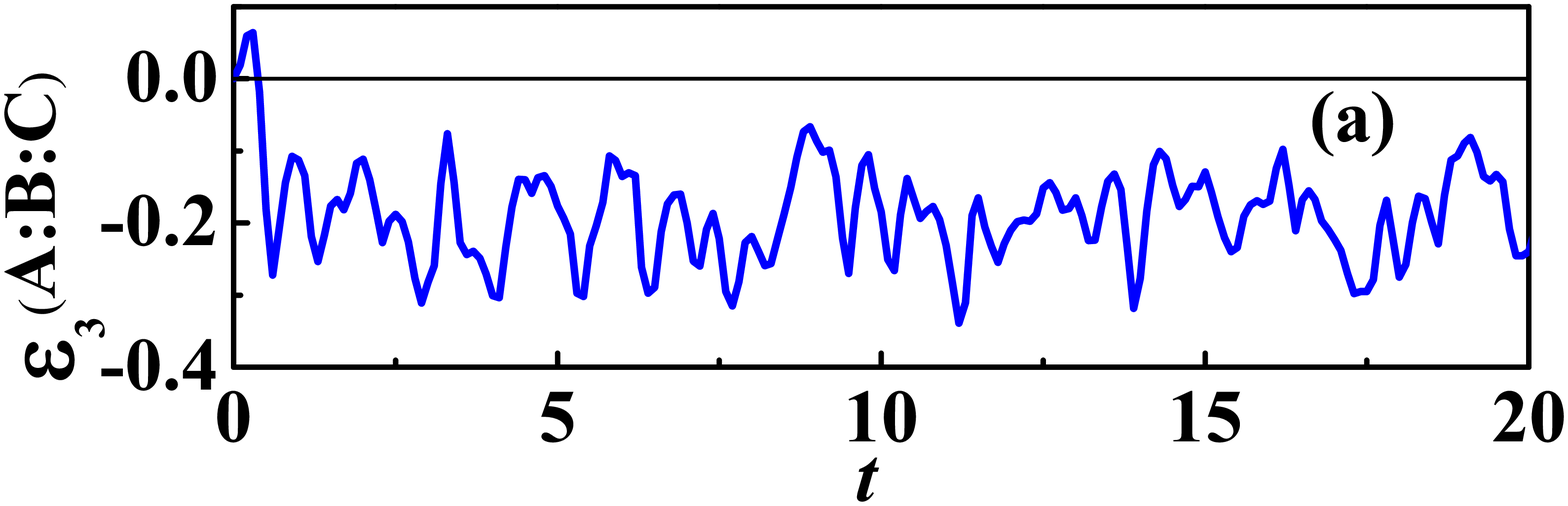}
\centering
\includegraphics[width=8.3cm]{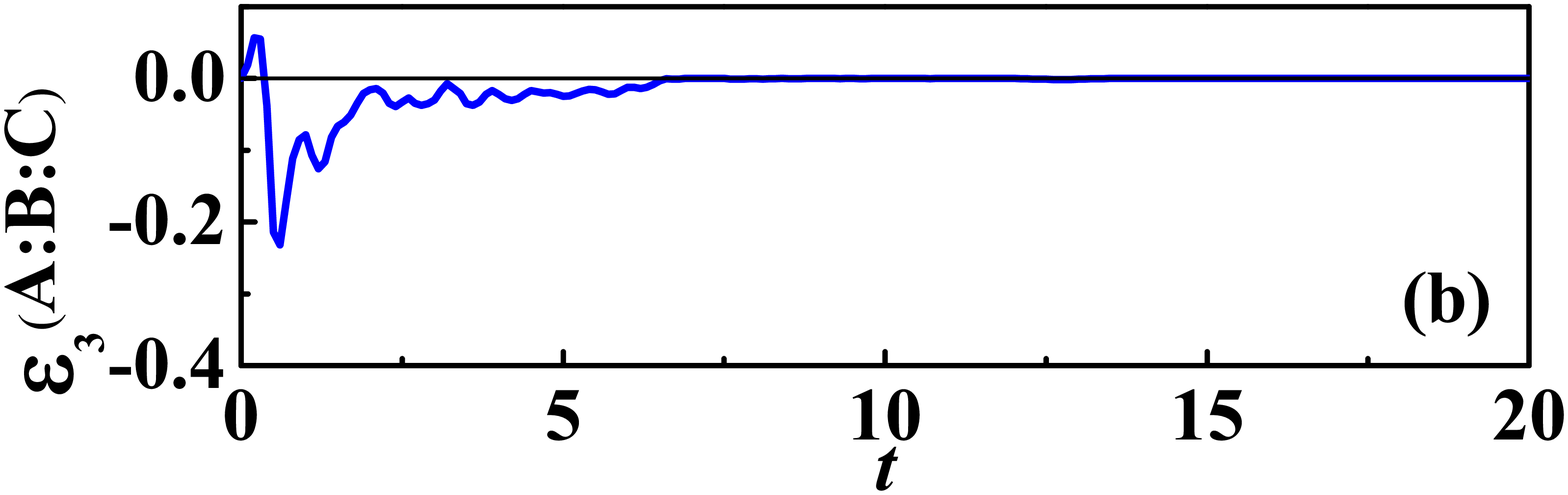}
\centering
\includegraphics[width=8.3cm]{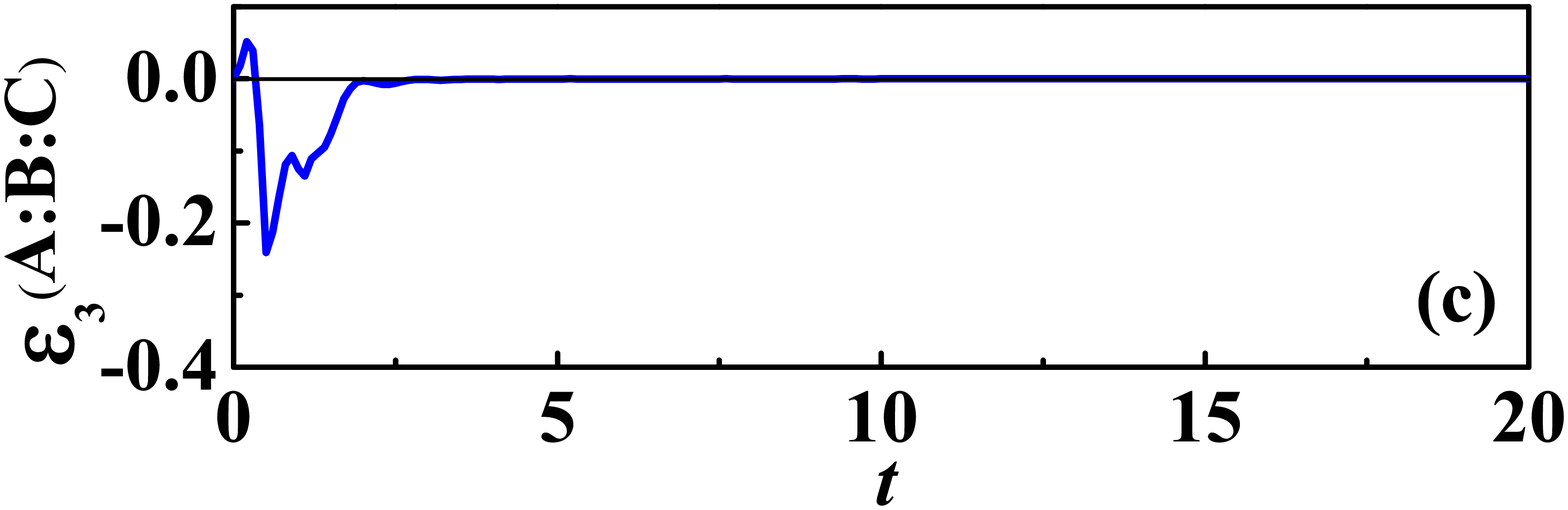}
\caption{TLN of $XXZ$ chain as a function of time for initial N\'{E}EL state, (a) in the absence of bath ($\Gamma  = 0$), (b) $L = \sigma ^ -$, and (c) $L = \sigma ^ z$. All the parameters are the same as those in Fig. 2.  }
\end{figure}

Next, we consider initial state $\left| {00...00} \right\rangle$. Fig. 4 shows the time evolutions of TMI and TLN in the absence of baths ($\Gamma  = 0$), while Fig. 5 shows them in the presence of baths respectively. As shown in Fig. 4(a), TMI is non-negative without baths for this initial state, implying that the information is not scrambled. Similar result for TLN is shown in Fig. 4(b). The reason why the information is not scrambled for this initial state in unitary case is that there is only one excitation for this initial state and thus there are few quasi-particles \cite{88}, which confines the dynamics and hence constrains the amount of entanglement that can emerge. Accordingly, quantum information is stored mostly in bipartite partitions and can not spread properly over many degrees of freedom. In contrast, in Fig. 5(a), TMI in the presence of dephasing baths can become slightly negative, which implies that the total information delocalization takes place. In Fig. 5(b), TLN in the presence of dephasing baths can also become slightly negative, which means that quantum information scrambling also occurs. These phenomena can be understood as that the system-bath interaction destroys the quasi-particle and thus changes the localized dynamics to a delocalized one. On the other hand, after a long time evolution, TMI reaches a steady negative value as shown in Fig. 5(a), while TLN decays to zero as shown in Fig. 5(b). It indicates that quantum information is totally lost and quantum information scrambling can only exist for a short time, while there is some remaining classical information at last and the total information delocalization still exists after a long time evolution.  Completely different from the results for channel $L = \sigma ^z$, both TMI and TLN still stay non-negative for channel $L = \sigma ^ -$ shown in Figs. 5(c) and 5(d) which means that both the total information delocalization and quantum information scrambling do not occur. It can be understood as the different sizes of the effective subspace for these two different baths. Similar to the case for initial N\'{E}EL state, for dephasing bath $L = \sigma ^z$, as mentioned above the total number of excitations is conserved, and thus the effective Hilbert subspace for quantum information is not decreased. However, different from the case for initial N\'{E}EL state, when initial state is $\left| {00...00} \right\rangle$, the total number of excitations for both ancillary qubit A and the system is gradually decreased from 1 to 0 for $L = \sigma ^ -$, and thus the effective Hilbert subspace for quantum information is always decreased. Hence, the amount of entanglement that can emerge is severely limited for $L = \sigma ^ -$. In a word, quantum information scrambling can occur for $L = \sigma ^z$, while for this initial state it can not occur in the case of $L = \sigma ^ -$.

\begin{figure}[htbp]
\centering
\includegraphics[width=8.3cm]{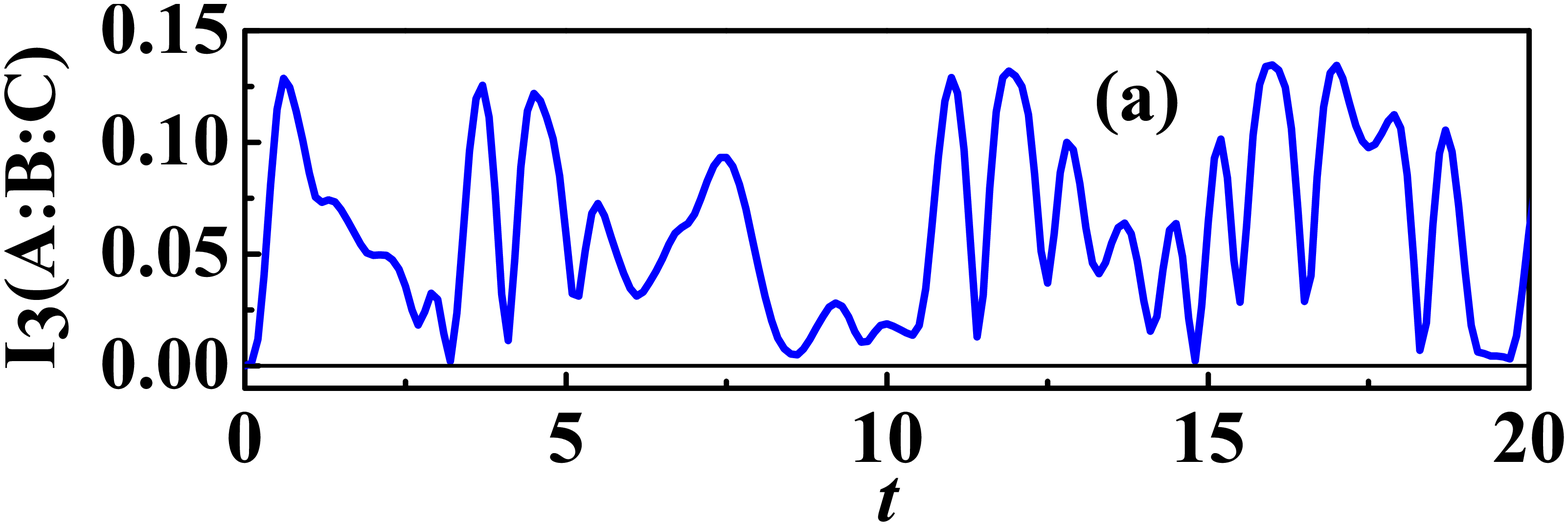}
\includegraphics[width=8.3cm]{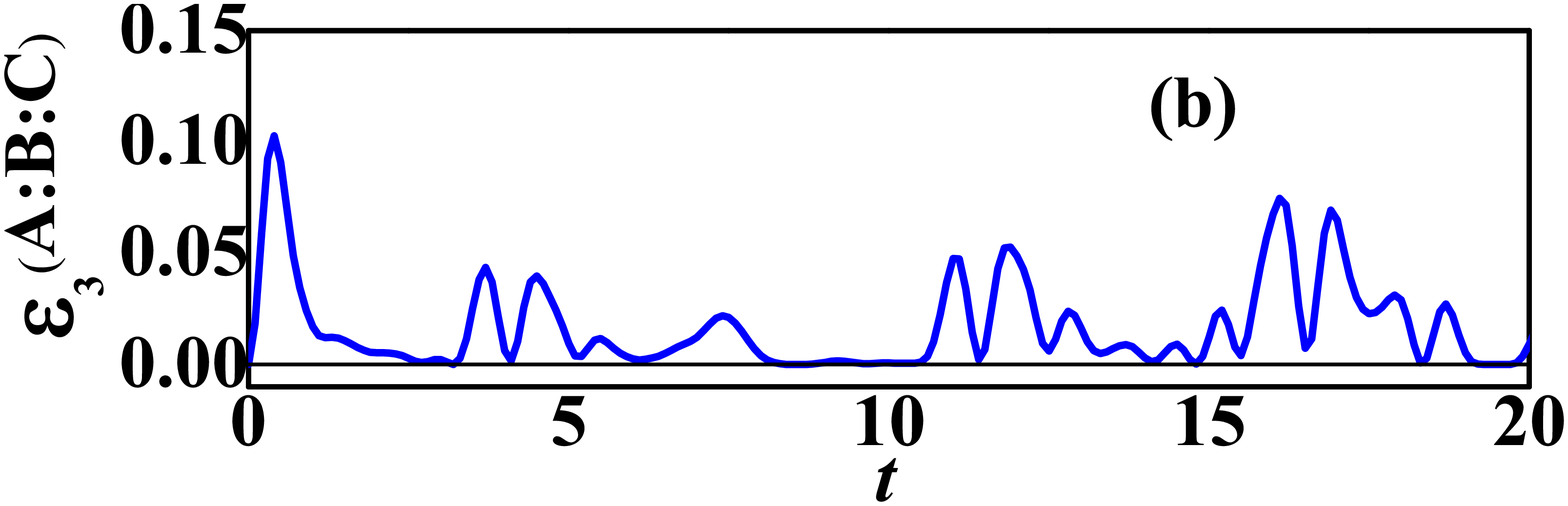}
\caption{TMI and TLN of $XXZ$ chain as functions of time for initial state $\left| {00...00} \right\rangle$ in the unitary case ($\Gamma  = 0$), (a) for TMI, and (b) for TLN. Here we choose $N = 7$, $n = 1$.}
\end{figure}

\begin{figure}[htbp]
\centering
\includegraphics[width=8.3cm]{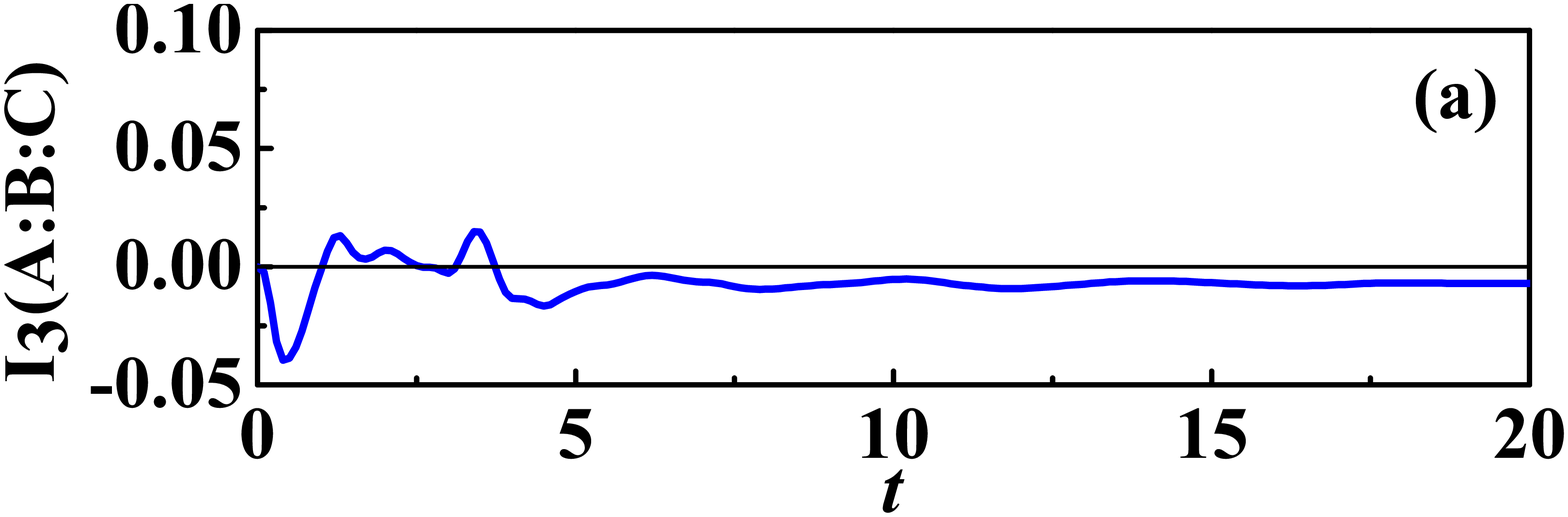}
\includegraphics[width=8.3cm]{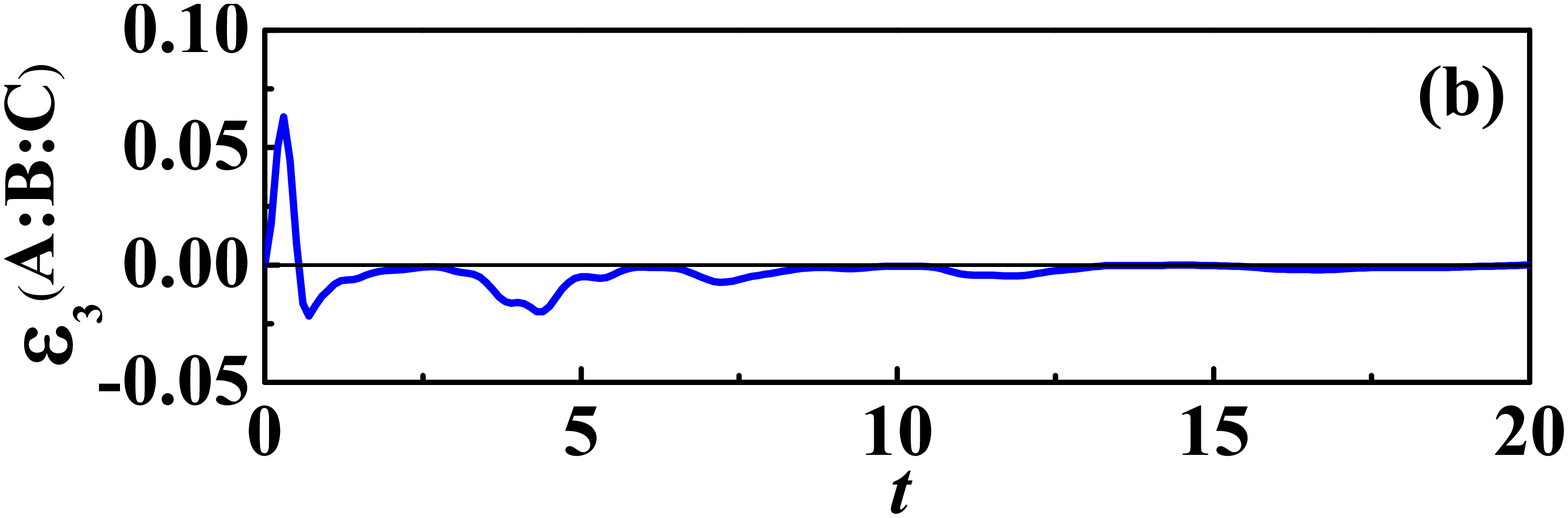}
\includegraphics[width=8.3cm]{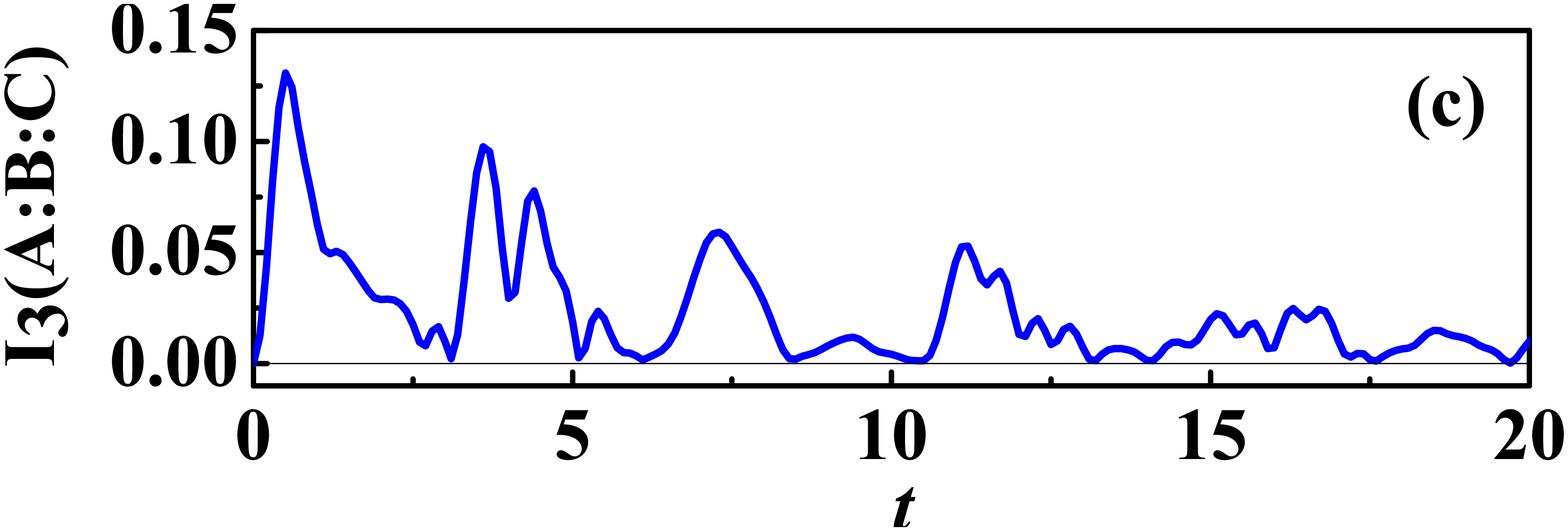}
\includegraphics[width=8.3cm]{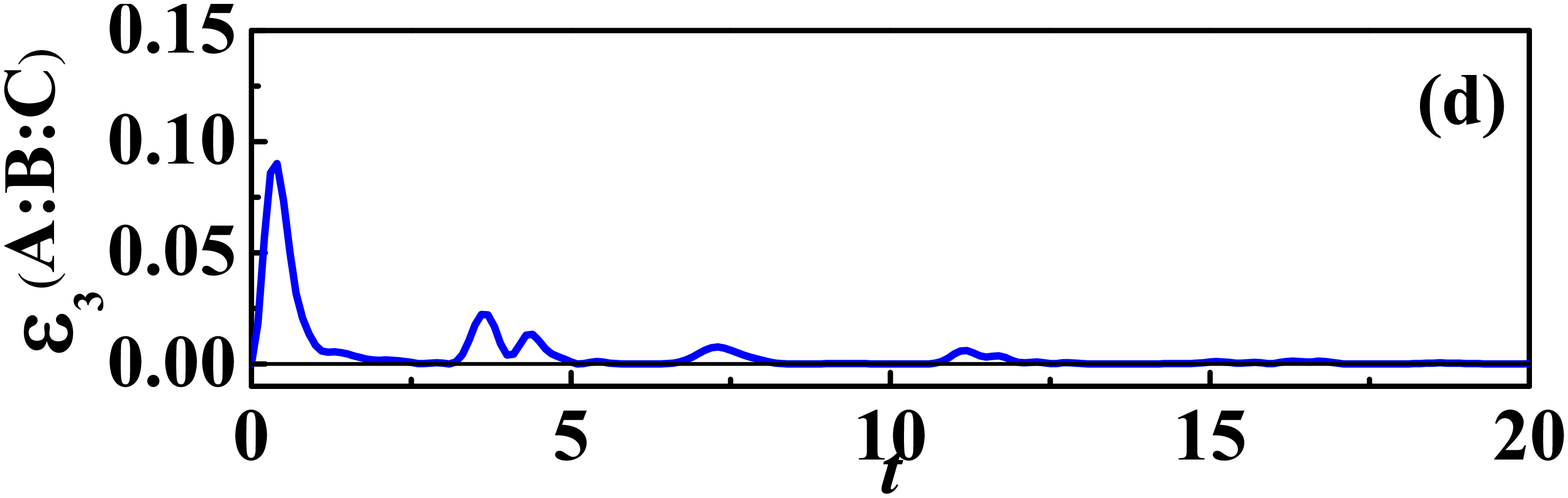}
\caption{TMI and TLN of $XXZ$ chain as functions of time for initial state $\left| {00...00} \right\rangle$, (a) and (b) for TMI and TLN in the case of  $L = \sigma ^z$ respectively; (c) and (d) for TMI and TLN in the case of  $L = \sigma ^-$ respectively. Here $N = 7$, $n = 1$ and  the other parameters are the same as those in Fig. 2.}
\end{figure}

\subsection{Effects of non-Markovianity on information scrambling }

In the following, we will investigate the effects of non-Markovianity on quantum information scrambling.  Firstly, we consider initial N\'{E}EL state and plot the time evolution of TMI for $L = \sigma ^z$ and different $\gamma$ in Fig. 6, (a) for $\gamma  = 1$, (b) for $\gamma  = 2$, and (c) for $\gamma  \to \infty$ respectively. As shown in Fig. 6, the maximum absolute value of the negative value of TMI decreases with the increase of $\gamma$, which means that it is  smaller for Markovian baths than that for non-Markovian baths. On the other hand, the time interval before TMI reaches its steady value becomes shorter with the increase of $\gamma$, in another word it is shorter for Markovian baths than that for non-Markovian baths.

It is known that $\gamma$ indicates the memory effect of the environment, and the smaller the $\gamma$, the longer the environmental memory time. While when $\gamma$ is small enough, non-Markovian properties can be observed. It has been shown that non-Markovianity due to the information backflow can be traced back to the establishment of correlations between system and environment as well as the changes in the state of the environment \cite{89,90,91}. In Markovian case, information of the system flows completely  into the environment.  While in non-Markovian case, information flow from the system partially preserved during the transient period  in the correlation between the system and the environment as well as in the environment, and will subsequently flow back to the  system.  It can be seen from Fig. 6 that with the decreasing of $\gamma$, the oscillation lasts longer time and decays more slowly for TMI. Thus the total information delocalization lasts for a longer time in the non-Markovian case.

It is noticed that the steady value of TMI achieved finally is independent of $\gamma$, which implies that the remaining classical information for different non-Markovianity is  the same. This is reasonable because in most cases non-Markovianity  affects the transient period of the system dynamics, rather than its steady state. Fig. 7 shows the time evolution of TLN which is similar to that of TMI. Clearly, the presence of the baths will suppress quantum information scrambling. However, it can be seen from Fig. 7 that with the increase of non-Markovianity the maximum absolute value of the negative value of TLN increases and it takes more time for TLN to decay to zero which means that non-Markovianity can enhance quantum information scrambling. This suggests that baths with memory will be beneficial to  the emergence of quantum information scrambling.

\begin{figure}[h!]
\centering
\includegraphics[width=8.3cm]{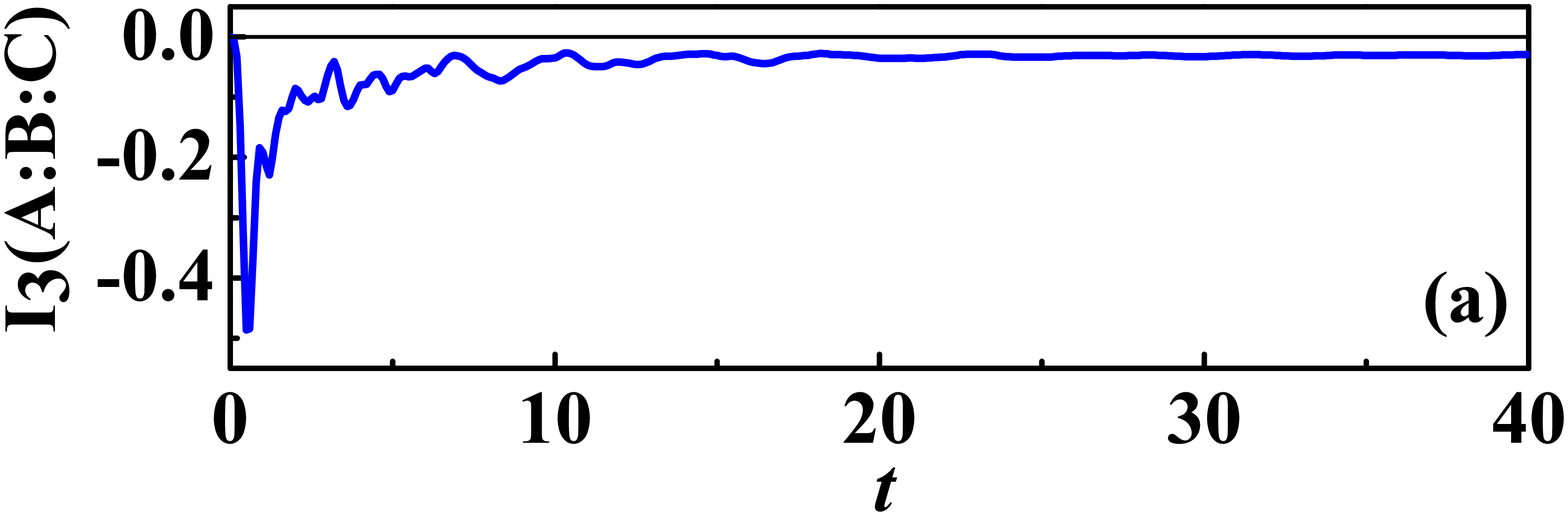}
\includegraphics[width=8.3cm]{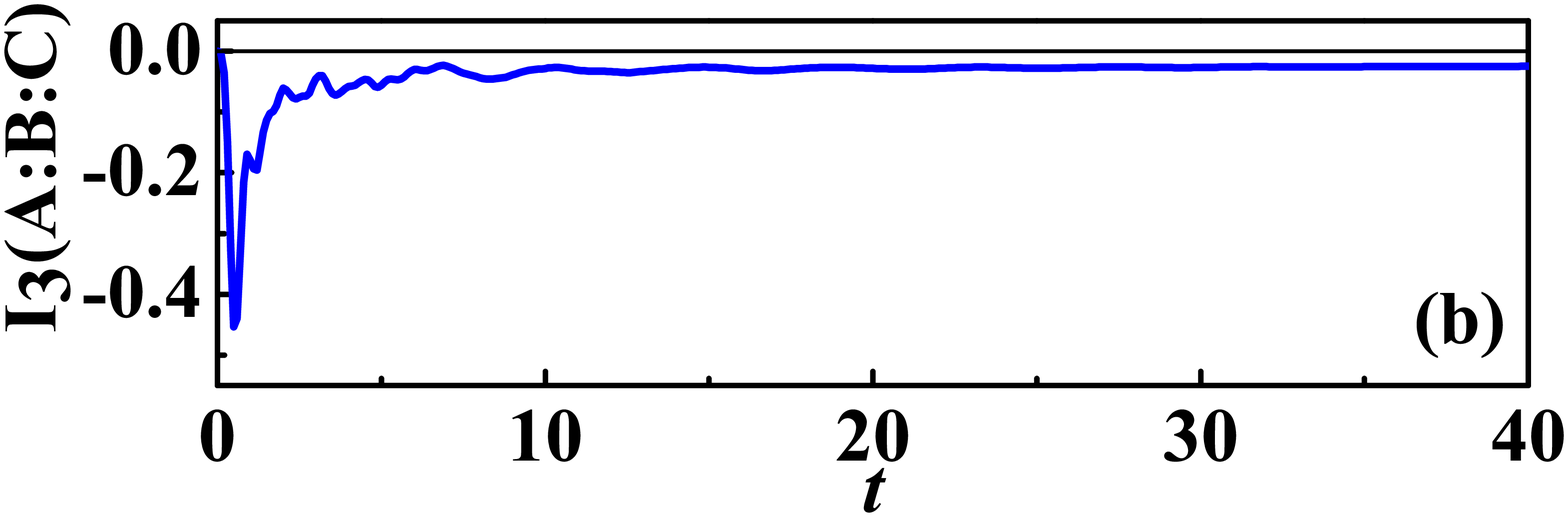}
\includegraphics[width=8.3cm]{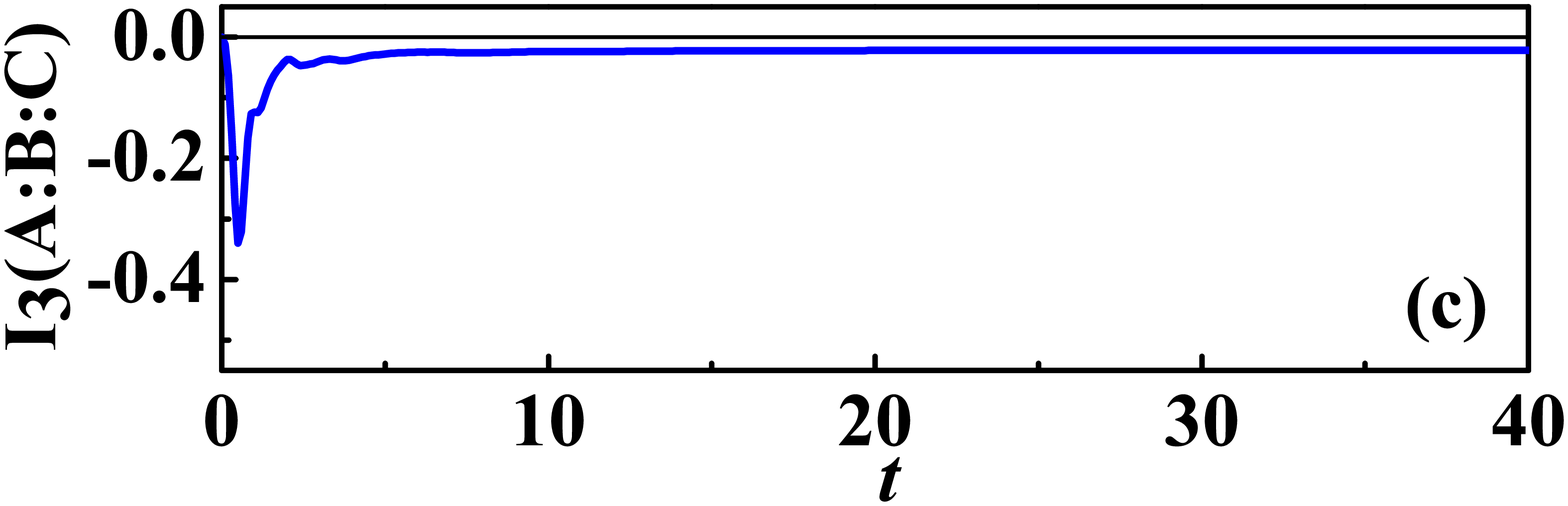}
\caption{TMI of $XXZ$ chain versus time $t$ in the case of  $L = \sigma ^ z$ for initial N\'{E}EL state and different $\gamma$, (a) $\gamma  = 1$, (b) $\gamma  = 2$, and (c) $\gamma  \to \infty$. The other parameters are $\Gamma  = 0.5$, $N = 6$, $n = 2$.}
\end{figure}

\begin{figure}[h!]
\centering
\includegraphics[width=8.3cm]{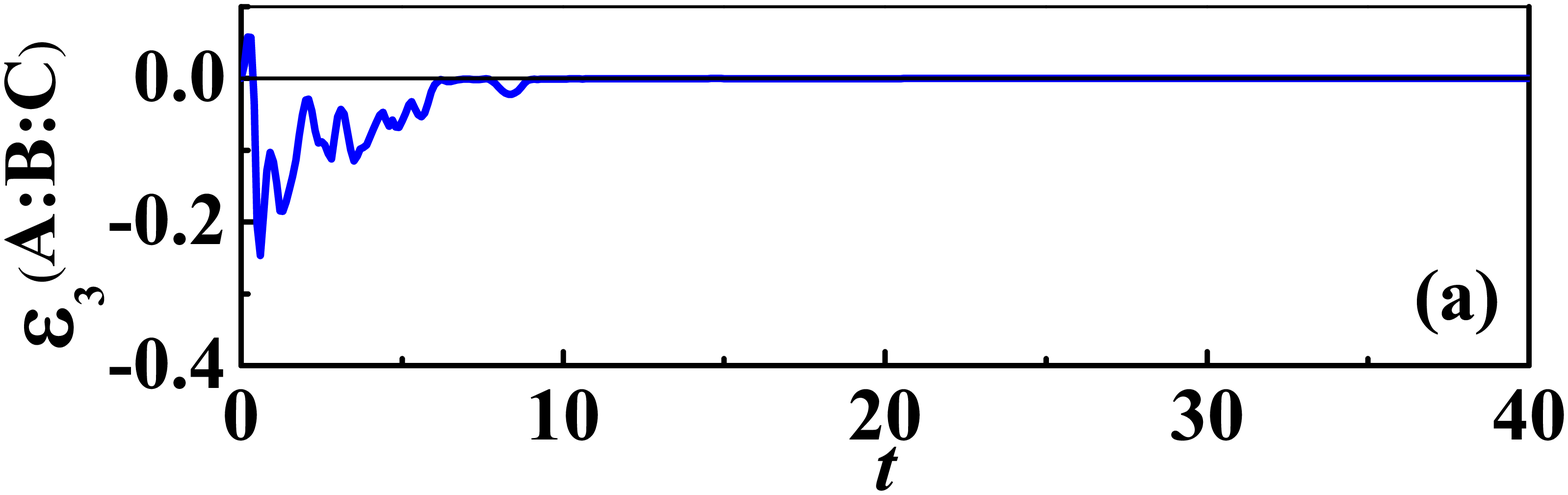}
\includegraphics[width=8.3cm]{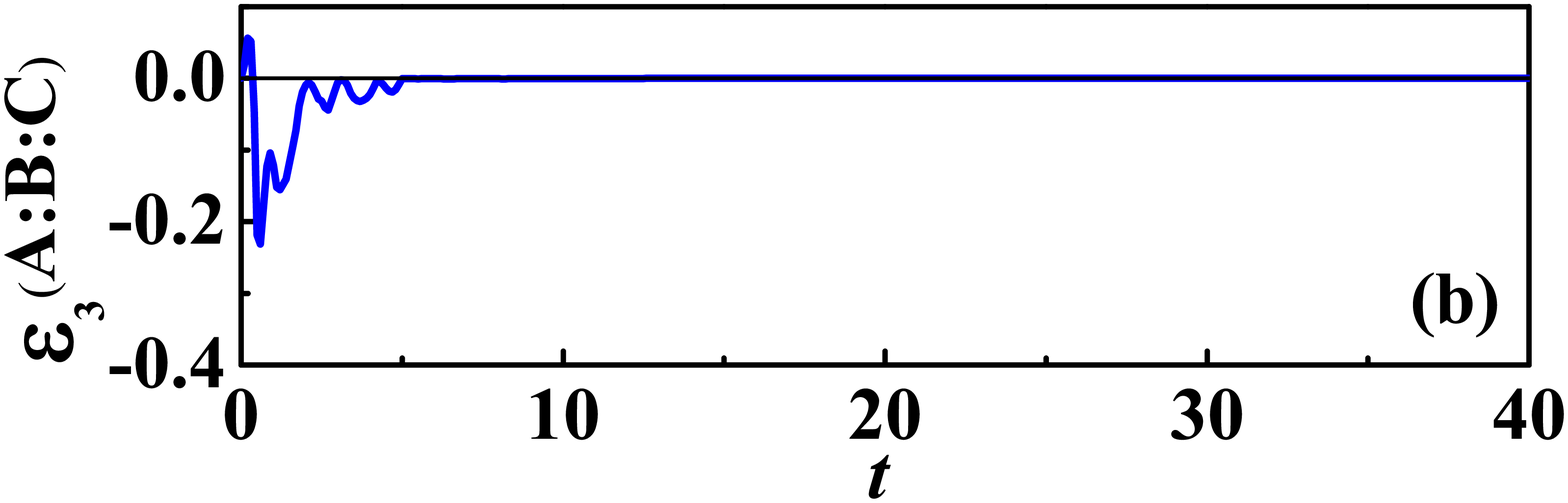}
\includegraphics[width=8.3cm]{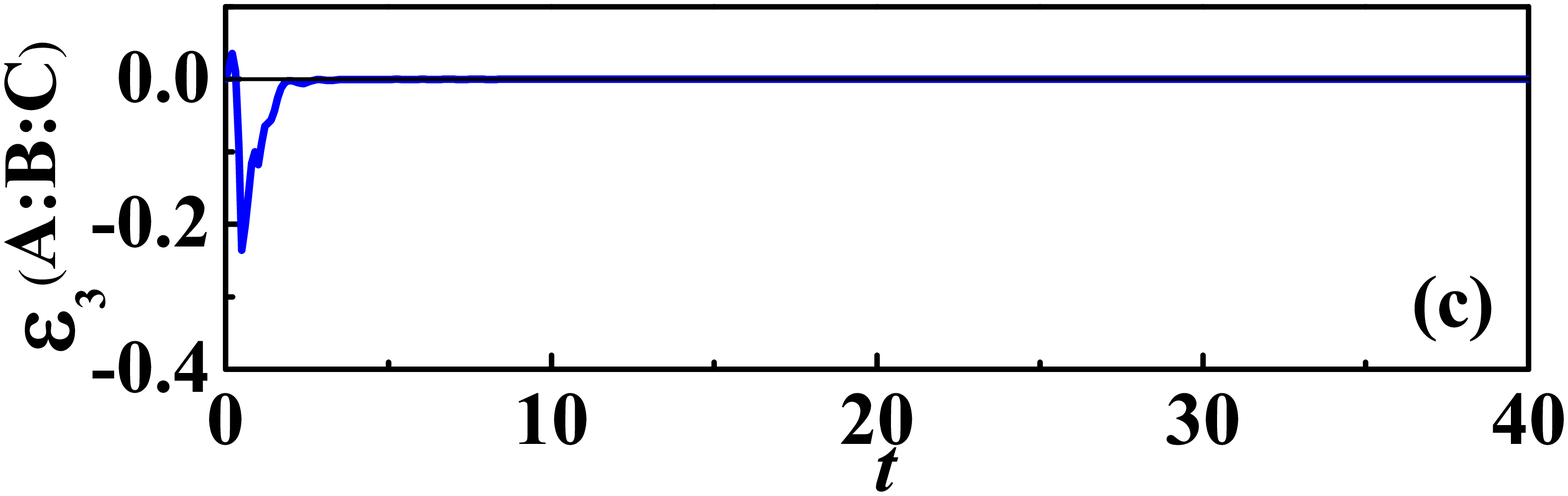}
\caption{TLN of $XXZ$ chain versus time $t$ in the case of  $L = \sigma ^ z$ for initial N\'{E}EL state and different $\gamma$, (a) $\gamma  = 1$, (b) $\gamma  = 2$, and (c) $\gamma  \to \infty$. The other parameters are the same as those in Fig. 6.}
\end{figure}

Then, we consider $L = \sigma ^ -$.  Again the results show that the memory effects of the baths can be helpful for the total information delocalization and quantum information scrambling, which are qualitatively the same as those for $L = \sigma ^ z$ (see Appendix B).


Next, we consider the effect of non-Markovianity on quantum information scrambling for initial state $\left|{00...00}\right\rangle$.  The results indicate that baths with memory can also enhance information delocalization for $L = \sigma ^ z$  which are qualitatively the same as those for initial N\'{E}EL state (see Appendix B). For $L = \sigma ^ -$, both information delocalization and quantum information scrambling can not occur whether in the Markovian or non-Markovian regimes.
Furthermore, we consider the influences of size of subsystem C on the total information delocalization and quantum information scrambling in the presence of baths. We find that with the increase in size of C information scrambling can last longer time for $XXZ$  chain (see Appendix B).
In addition, we investigate the effect of $\Gamma$ on TMI and TLN. We find that the maximum absolute value of the negative value for TLN as well as the time duration before it reaches zero decrease with increasing $\Gamma$, which implies that a stronger system-bath interaction corresponds to less quantum information scrambling. The effect of $\Gamma$ on TMI is almost the same as that on TLN, hence the stronger the system-bath interaction is, the
weaker the total information delocalizaion becomes.
For the $XX$ chain ($\Delta  = 0$), which is a non-interacting model, the results for TMI and TLN are almost the same as those for interacting model ($XXZ$ chain), and only slightly different (see Appendix C).




\section{CONCLUSION}
In this paper, we have studied the total information delocalization and quantum information scrambling by using tripartite mutual information, and tripartite logarithmic negativity. The model we have considered is a spin chain with two ends interacting with two separate baths and  we have used a non-Markovian quantum state diffusion equation approach to obtain the time evolutions of TMI and TLN. We have considered two types of system-bath interactions, i.e., dephasing channel and dissipation channel as well as two types of initial system states, N\'{E}EL state and $\left| {00...00} \right\rangle$. Interestingly, it has been found that TMI can still be negative when there is no entanglement at all which means that negative TMI might not be a suitable quantifier of quantum information scrambling for open quantum system any more, but negative TLN is an appropriate one. By comparing the dynamics of TLN with TMI, we can distinguish the quantum information scrambling from the total information delocalization in open quantum system.  Our results have shown that generally the existence of baths suppresses both the total information delocalization and quantum information scrambling in a long time. However, in some cases environment can play a beneficial role, for example, for initial state $\left| {00...00} \right\rangle$, information is not scrambled in the absence of baths, while the total information delocalization and quantum information scrambling can occur in the early period in the presence of dephasing baths. These phenomena can be understood as that the system-bath interaction destroys the quasi-particle and thus change the localized dynamics to a delocalized one. More importantly it has  been found that non-Markovianity plays an beneficial role in both the total information delocalization and quantum information scrambling.
 Moreover, we have considered the influences of size of subsystem C on the total information delocalization and quantum information scrambling in the presence of baths. It has been found that the larger the size of subsystem C is, the longer time quantum information scrambling lasts for both $XXZ$ chain and $XX$ chain.

\begin{acknowledgments}
This work was supported by the National Natural Science Foundation of China (Grant Nos.
11775019 and 11875086).
\end{acknowledgments}

\appendix
\section{DERIVATION OF THE NON-MARKOVIAN MASTER EQUATION}

The derivation of the non-Markovian master equation by the QSD equation  is based on non-Markovian stochastic quantum trajectories. The basic idea is that the total wave function $\left| {\Psi _{\rm{tot}} (t)}
\right\rangle$ is projected into the coherent state of the bath mode $\left| z \right\rangle$, and  $\left| {\Psi _{z^* } (t)} \right\rangle  = \left\langle {{z^* }}\mathrel{\left | {\vphantom {{z^* } {\Psi _{\rm{tot}} (t)}}}\right. \kern-\nulldelimiterspace}{{\Psi _{\rm{tot}} (t)}} \right\rangle$ which is known as stochastic quantum trajectory. It obeys a linear  QSD equation  \cite{49,80}
\begin{align}
&\frac{\partial }{{\partial t}}\left| {\Psi _{z^ *  } \left( t \right)} \right\rangle\nonumber\\
&= \left\{ { - iH_s  + \sum\limits_{j = 1,2} {\left[ {L_j z_{jt}^ *   - L_j^\dag  \bar O_j \left( {t,z^ *  } \right)} \right]} } \right\}\left| {\Psi _{z^ *  } \left( t \right)} \right\rangle,\label{A 1}
\end{align}
where $z_{jt}^*  =  - i\sum\limits_k {g_{jk} z_{jk}^* e^{i\omega _{jk} t} }$ is a Gaussian stochastic process, $O$ is an operator defined by $
\frac{\delta }{{\delta z_{js}^* }}\left| {\Psi _{z^* } (t)} \right\rangle  = O_j \left( {t,s,z_1^* ,z_2^* } \right)\left| {\Psi _{z^* } (t)} \right\rangle$, and $\bar O_j \left( {t,z^* } \right) = \int_0^t {\alpha _j \left( {t,s} \right)O_j } \left( {t,s,z_1^* ,z_2^* } \right)ds$. Assuming the bath is at zero temperature, the correlation function is $\alpha _j \left( {t,s} \right) = \sum\limits_k {\left| {g_{jk} } \right|} ^2 e^{ - i\omega _{jk} \left( {t - s} \right)}$, describing the effect of the bath and $M\left[ {z_{jt}^* z_{js}^{} } \right] = \alpha _j \left( {t,s} \right)$, where $M\left[  \cdot \right]$ is the ensemble average.

According to the consistency condition, the $O$ operator satisfies \cite{80}
\begin{align}
&\frac{\partial }{{\partial t}}O_1 \left( {t,s,z_1^ *  ,z_2^ *  } \right) =  - i\left[ {H_s ,O_1 \left( {t,s,z_1^ *  ,z_2^ *  } \right)} \right] \nonumber\\
&+ \left[ {\sum\limits_{j = 1,2} {\left( {L_j z_{jt}^ *   - L_j^\dag  \bar O_j \left( {t,z_1^ *  ,z_2^ *  } \right)} \right),O_1 \left( {t,s,z_1^ *  ,z_2^ *  } \right)} } \right]\nonumber\\
&- \sum\limits_{j = 1,2} {L_j^\dag  \frac{\delta }{{\delta z_{2s}^ *  }}\bar O_j \left( {t,s,z_1^ *  ,z_2^ *  } \right)},\label{A 2}
\end{align}
\begin{align}
&\frac{\partial }{{\partial t}}O_2 \left( {t,s,z_1^ *  ,z_2^ *  } \right) =  - i\left[ {H_s ,O_2 \left( {t,s,z_1^ *  ,z_2^ *  } \right)} \right] \nonumber\\
&+ \left[ {\sum\limits_{j = 1,2} {\left( {L_j z_{jt}^ *   - L_j^\dag  \bar O_j \left( {t,z_1^ *  ,z_2^ *  } \right)} \right),O_2 \left( {t,s,z_1^ *  ,z_2^ *  } \right)} } \right]\nonumber\\
&- \sum\limits_{j = 1,2} {L_j^\dag  \frac{\delta }{{\delta z_{1s}^ *  }}\bar O_j \left( {t,s,z_1^ *  ,z_2^ *  } \right)}.\label{A 3}
\end{align}

Instead of a direct numerical simulating the trajectories by the QSD equation above, we can analytically take the ensemble average to obtain a non-Markovian master equation. Based on Eqs. (A 2) and (A 3) the reduced density matrix of the system $\rho _s  = M\left[ {P_t } \right]$ can be obtained, where $P_t  = \left| {\Psi _{z^ *  } \left( t \right)} \right\rangle \left\langle {\Psi _z \left( t \right)} \right|$. Using Novikov's theorem \cite{92,93,94}, the general non-Markovian master equation can be derived \cite{95}
\begin{align}
\frac{\partial }{{\partial t}}&\rho _s  =  - i\left[ {H_s ,\rho _s } \right]\nonumber\\
&+ \sum\limits_{j = 1,2} {\left( {\left[ {L_j ,M\left[ {P_t \bar O_j^\dag  } \right]} \right]} \right.}- \left. {\left[ {L_j^\dag  ,\left. {M\left[ {\bar O_j P_t } \right]} \right]} \right.} \right).\label{A 4}
\end{align}
It is noticed that the above equation is still not a closed equation for $\rho _s$. Generally, the operator $\bar O_j$ contains noises $z_1^ *$, $z_2^ *$. When $\bar O_j \left( {t,z_1^ *  ,z_2^ *  } \right)$  is approximated by a noise-independent operator, i.e., $\bar O_j \left( {t,z_1^ *  ,z_2^ *  } \right) = \bar O_j \left( t \right)$, the evolution equation (5) is obtained.

 For the Ornstein-Uhlenbeck correlation, we have the relation
 \begin{align}
  \mathop {\dot\alpha _j }\limits^{} \left( {t,s} \right) =  - \gamma _j \alpha _j \left( {t,s} \right).
 \end{align}
 Using the above relation and Eqs. (A 2) and (A 3),  one can obtain Eqs. (6) and (7).

 \section{SUPPLEMENTAL NUMERICAL RESULTS FOR  $XXZ$ CHAIN}
 In Figs. 8 and 9 we plot the time evolutions of TMI and TLN for different  $\gamma$ and initial N\'{E}EL state in the case of $L = \sigma ^ -$ respectively. Again the results show that the memory effects of the baths can be helpful for the total information delocalization and quantum information scrambling, which are consistent with those in the case of $L = \sigma ^ z$  shown in Figs. 6 and 7.

 \begin{figure}[h!]
\centering
\includegraphics[width=8.3cm]{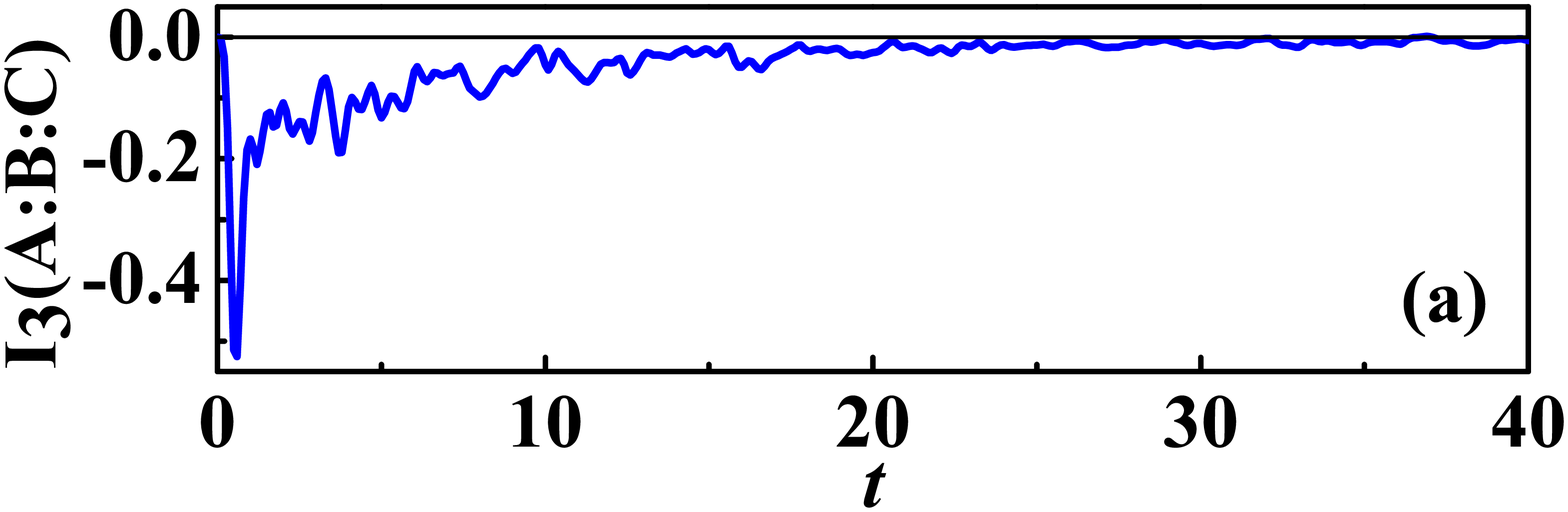}
\includegraphics[width=8.3cm]{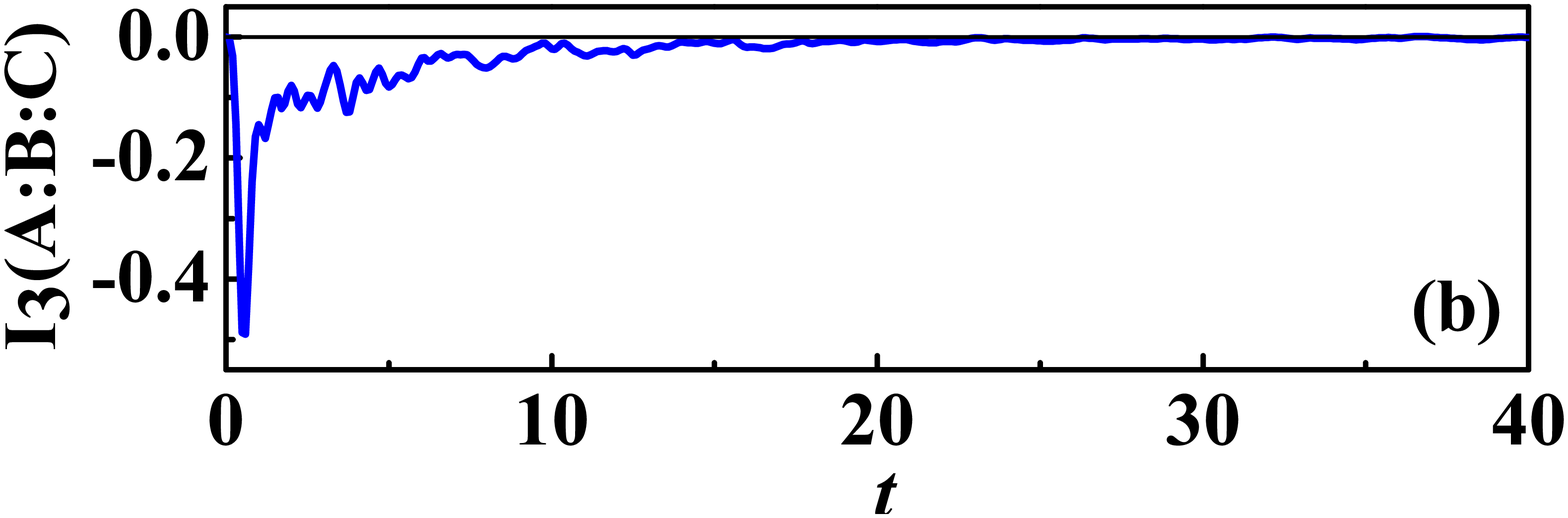}
\includegraphics[width=8.3cm]{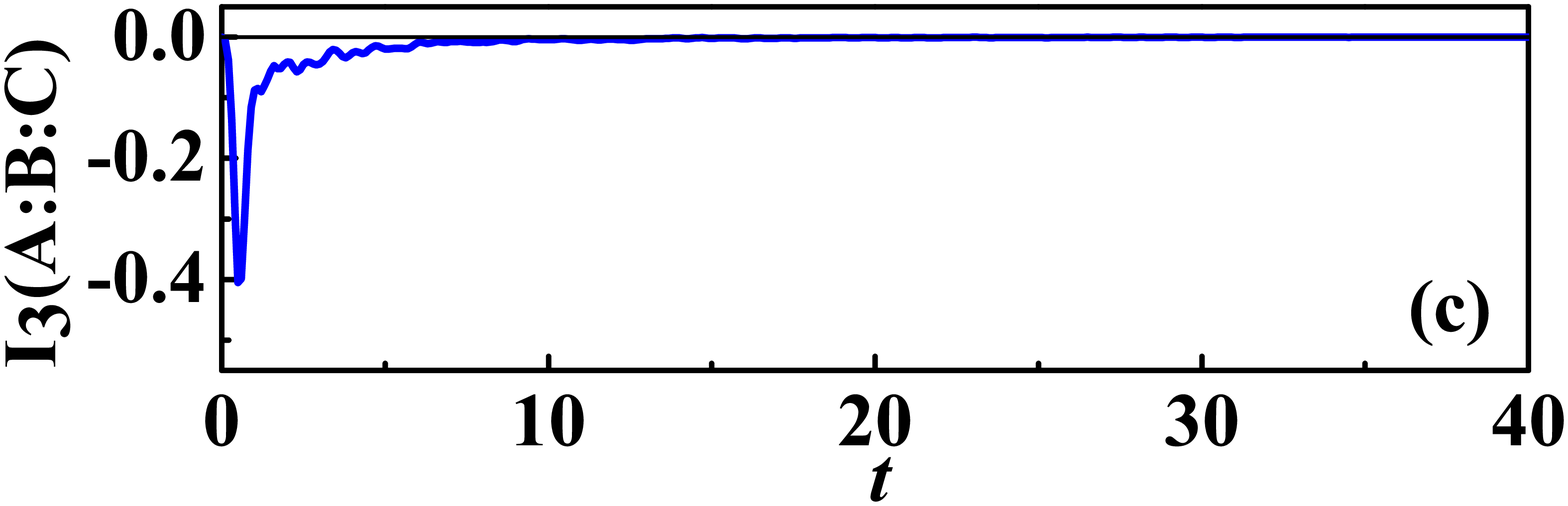}
\caption{TMI of $XXZ$ chain versus time $t$ in the case of  $L = \sigma ^ -$ for initial N\'{E}EL state and different $\gamma$, (a) $\gamma  = 1$, (b) $\gamma  = 2$, and (c) $\gamma  \to \infty$. The other parameters are the same as those in Fig. 6.}
\end{figure}

\begin{figure}[h!]
\centering
\includegraphics[width=8.3cm]{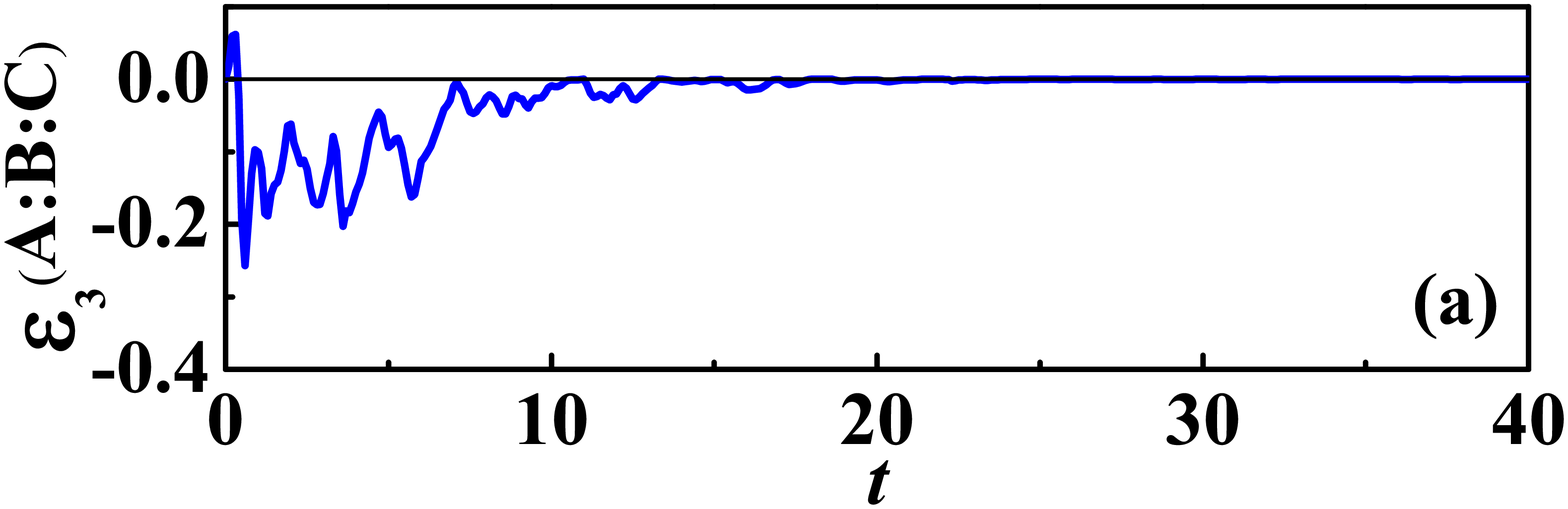}
\includegraphics[width=8.3cm]{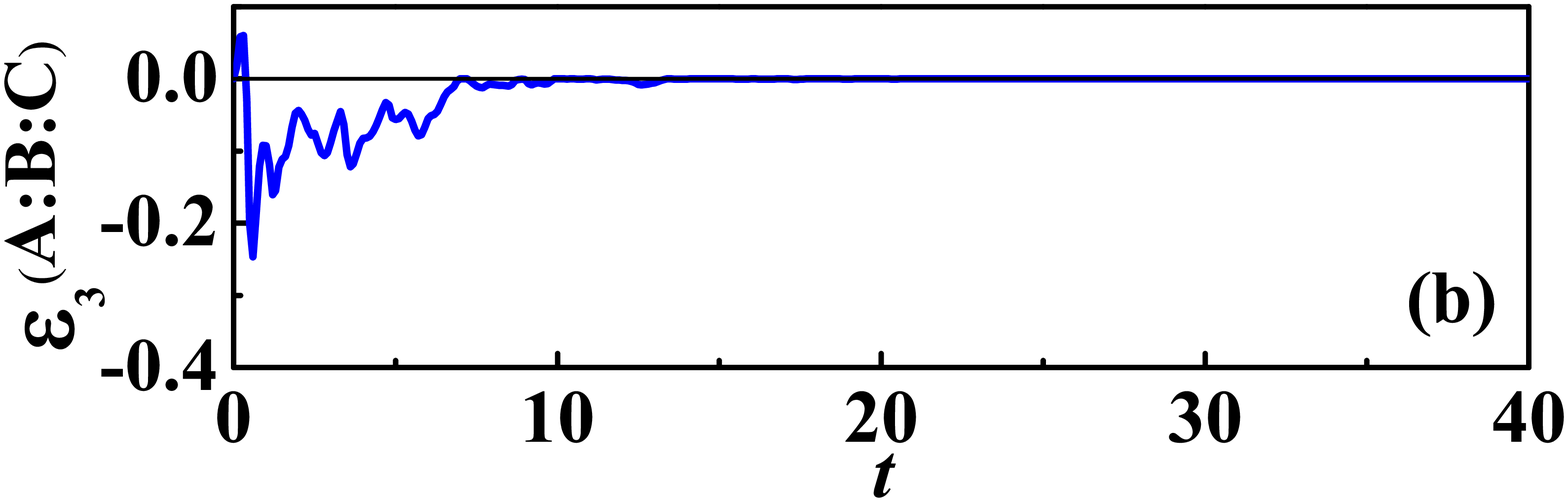}
\includegraphics[width=8.3cm]{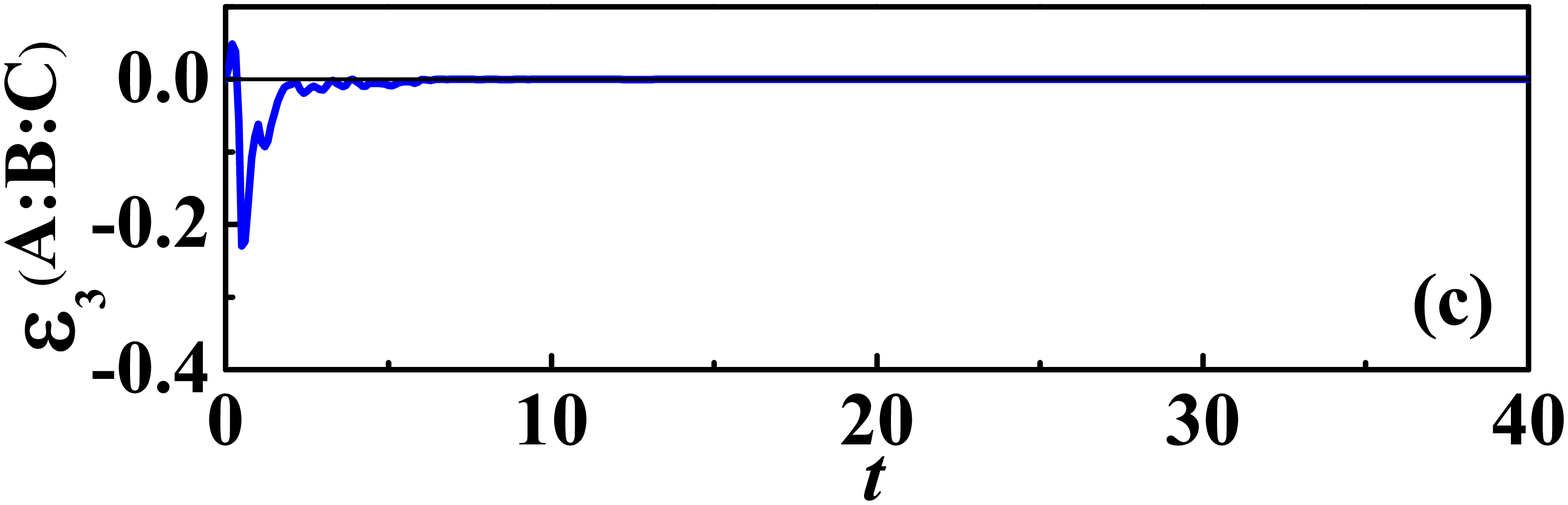}
\caption{TLN of $XXZ$ chain versus time $t$ in the case of  $L = \sigma ^ -$ for initial N\'{E}EL state and different $\gamma$, (a) $\gamma  = 1$, (b) $\gamma  = 2$, and (c) $\gamma  \to \infty$. The other parameters are the same as those in Fig. 6.}
\end{figure}

  Fig. 10 plots  the time evolutions of TMI and TLN for different $\gamma$ and initial state $\left| {00...00} \right\rangle$ in the case of $L = \sigma ^ z$. From Fig. 10, we can see that the time interval before TMI reaches its steady value or TLN reaches zero are decreasing with increasing $\gamma$, which are qualitatively the same as those for initial N\'{E}EL state.

 \begin{figure}[htbp]
\centering
\includegraphics[width=4cm]{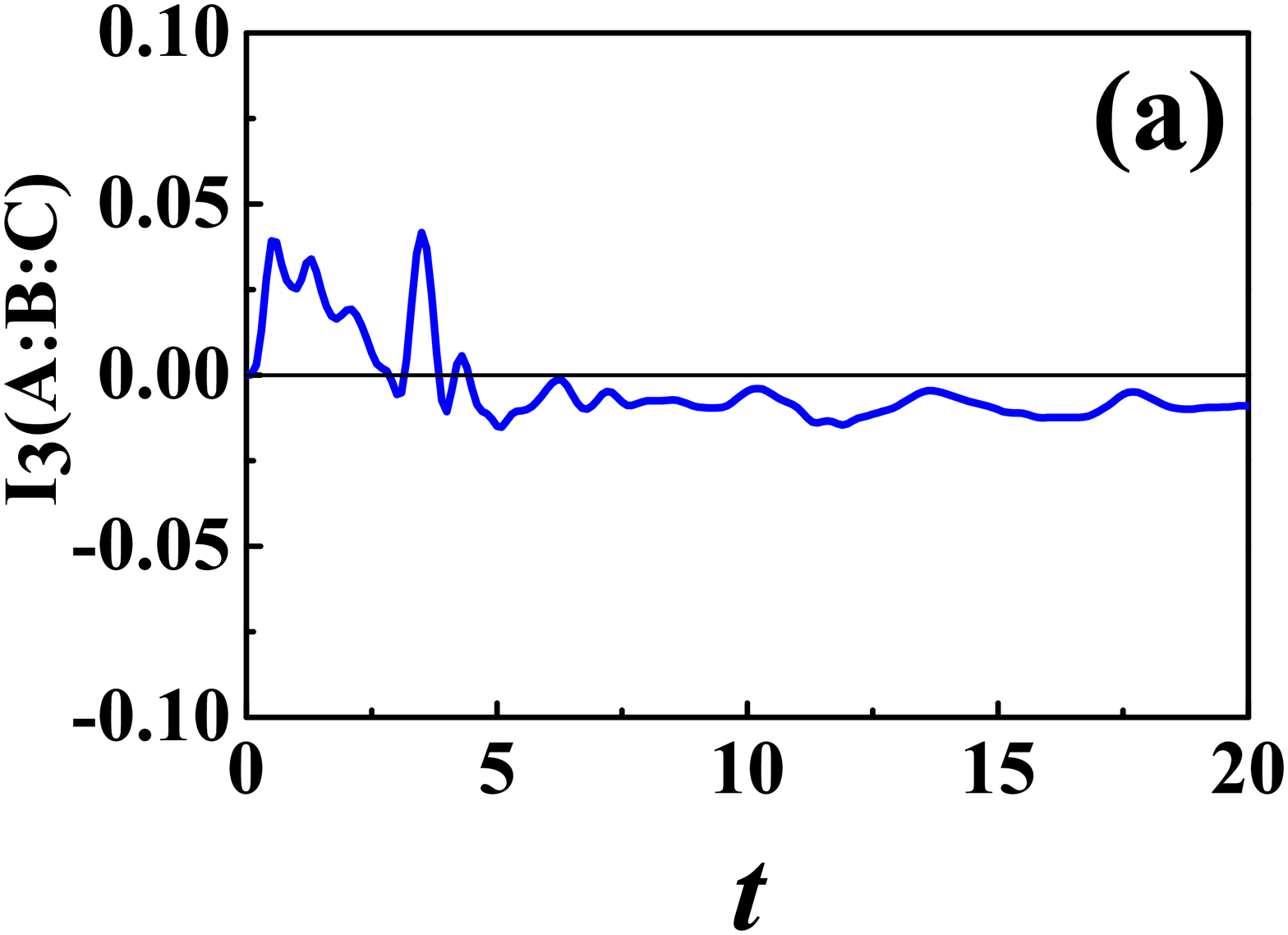}
\includegraphics[width=4cm]{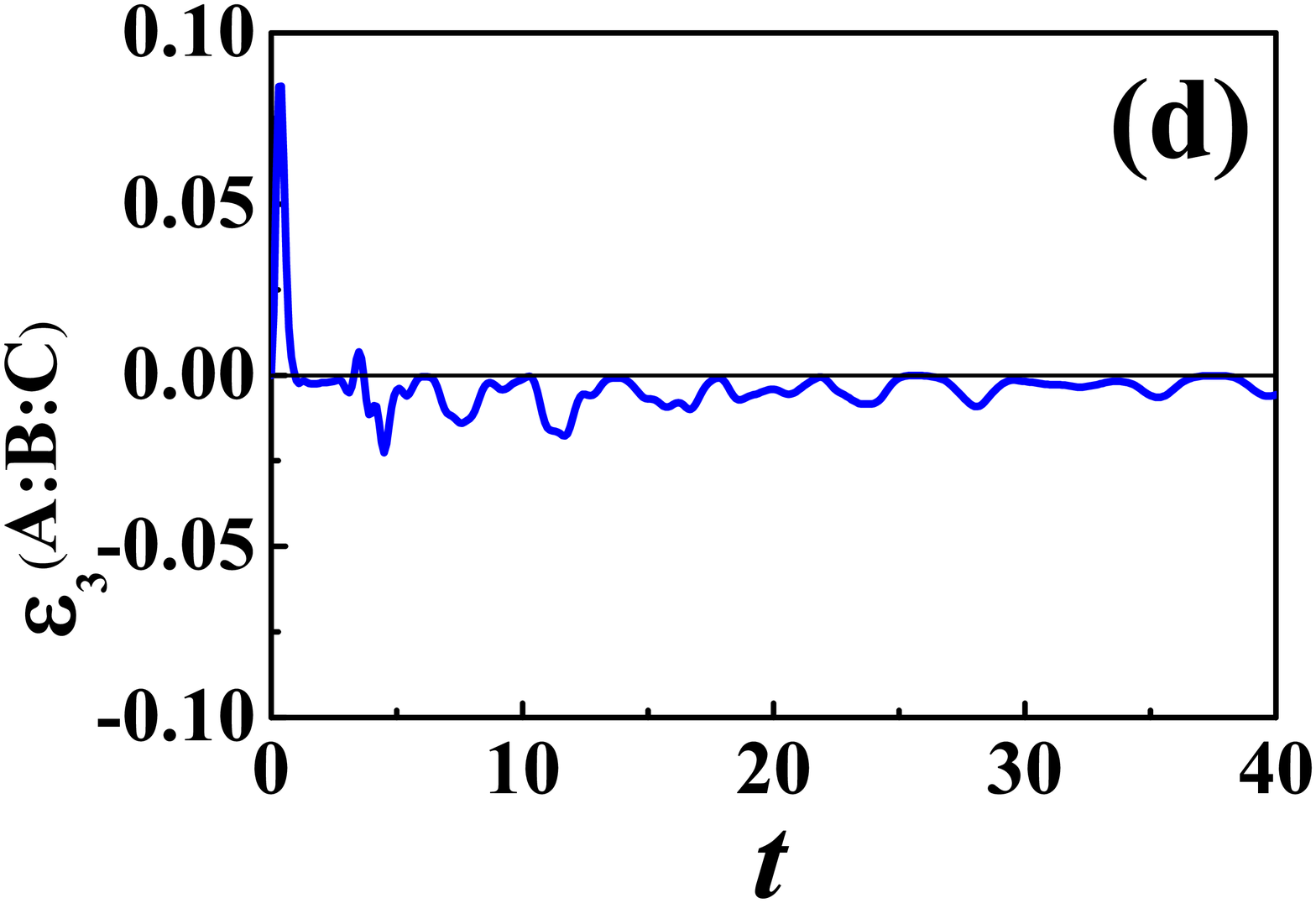}
\includegraphics[width=4cm]{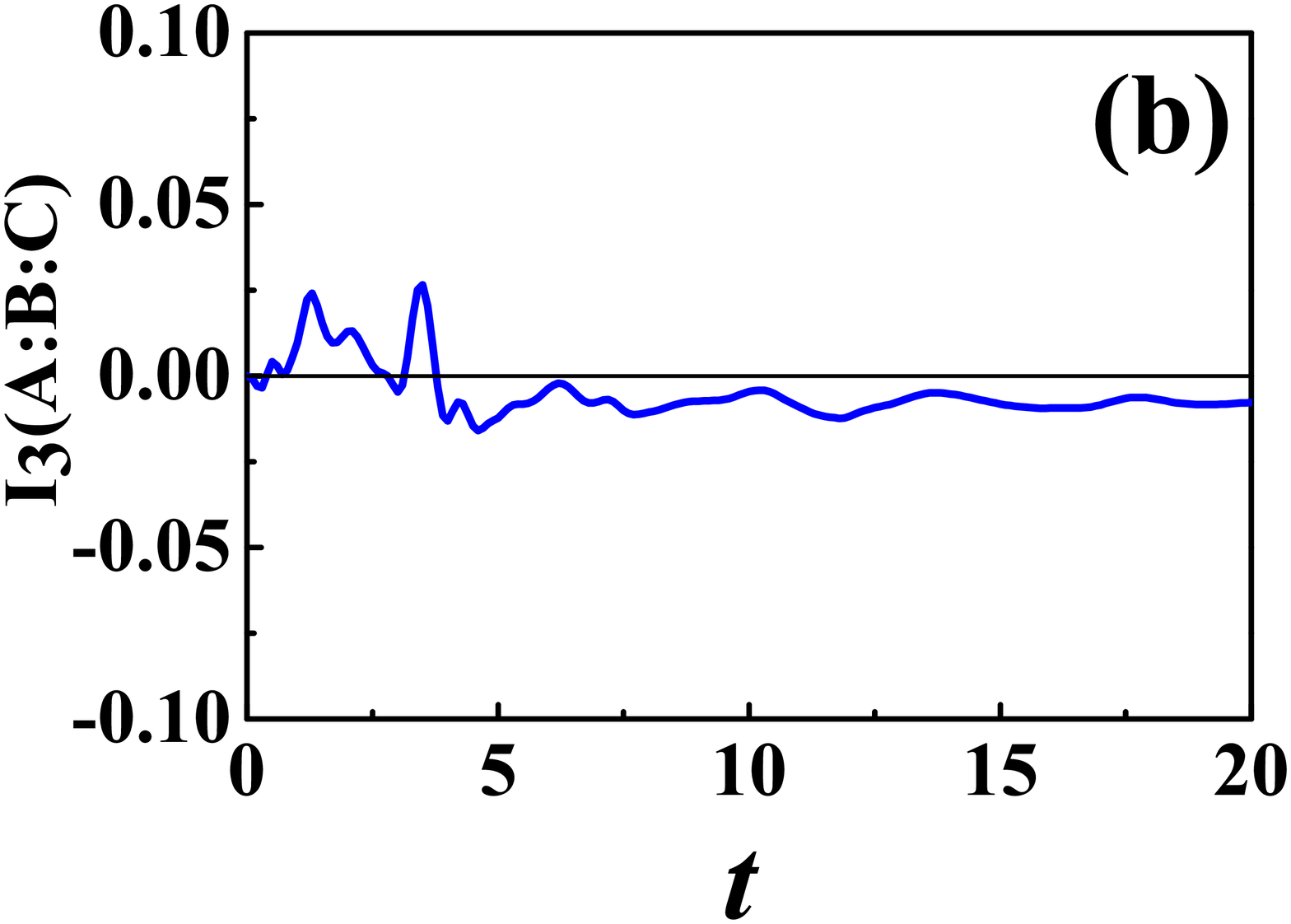}
\includegraphics[width=4cm]{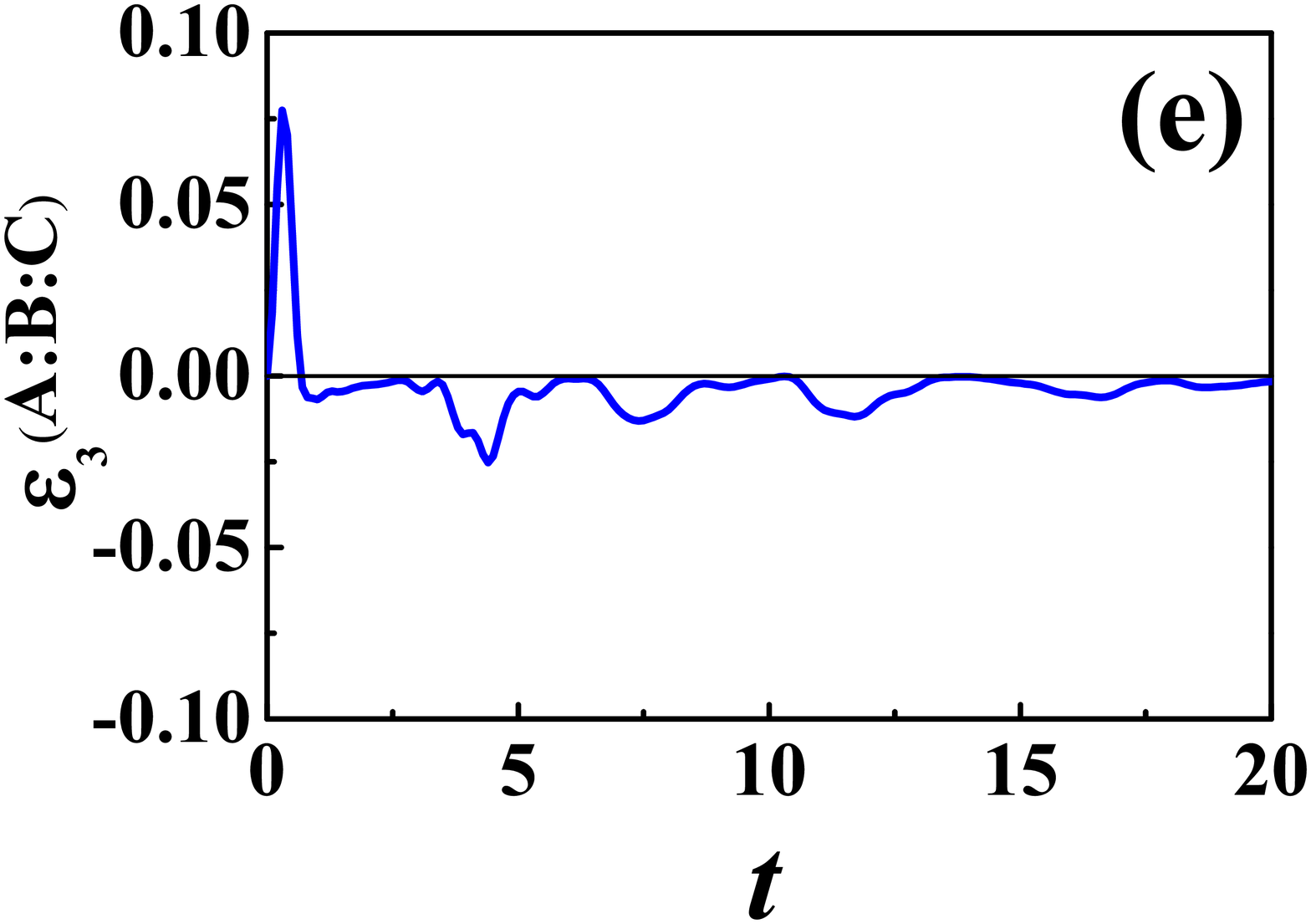}
\includegraphics[width=4cm]{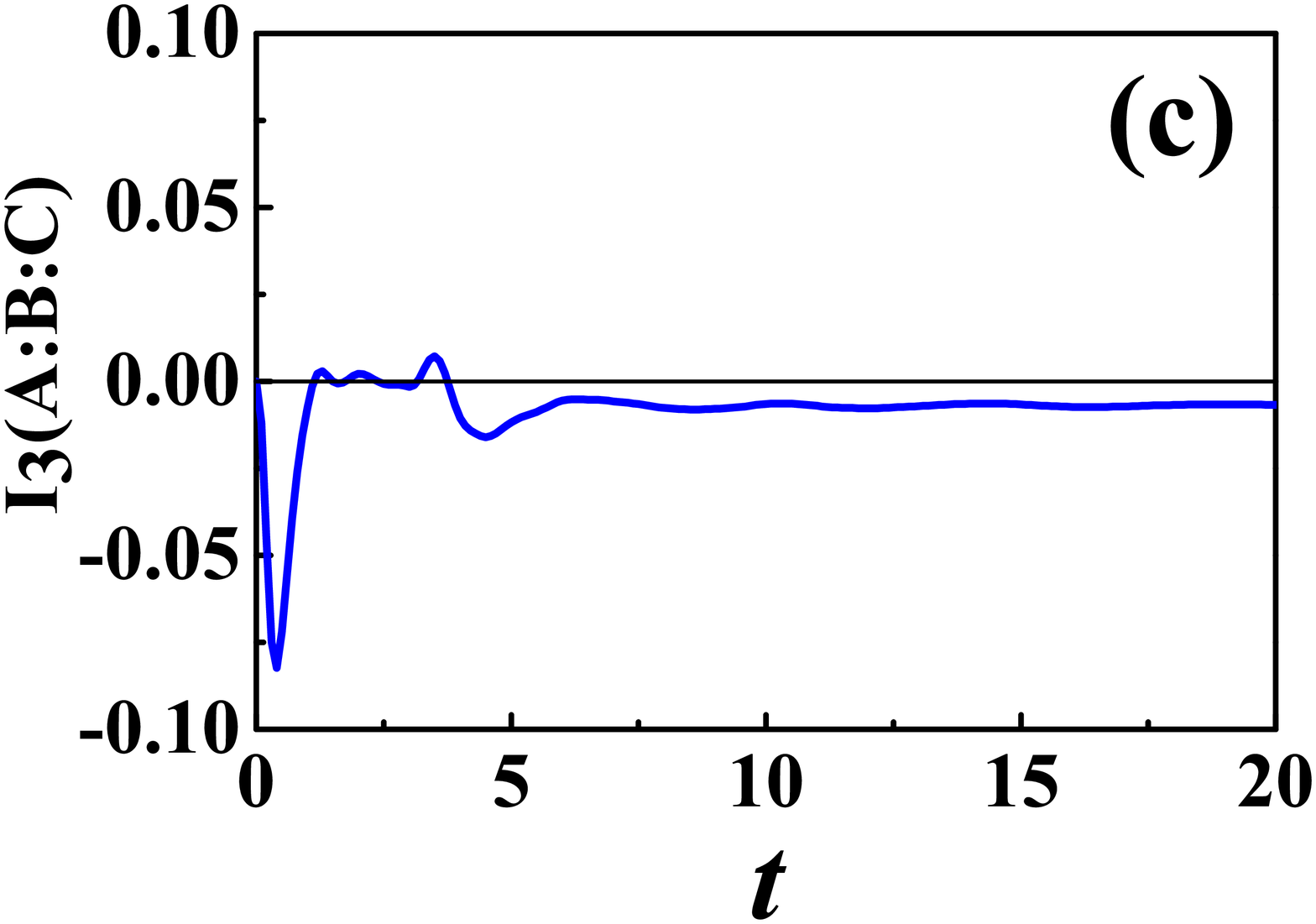}
\includegraphics[width=4cm]{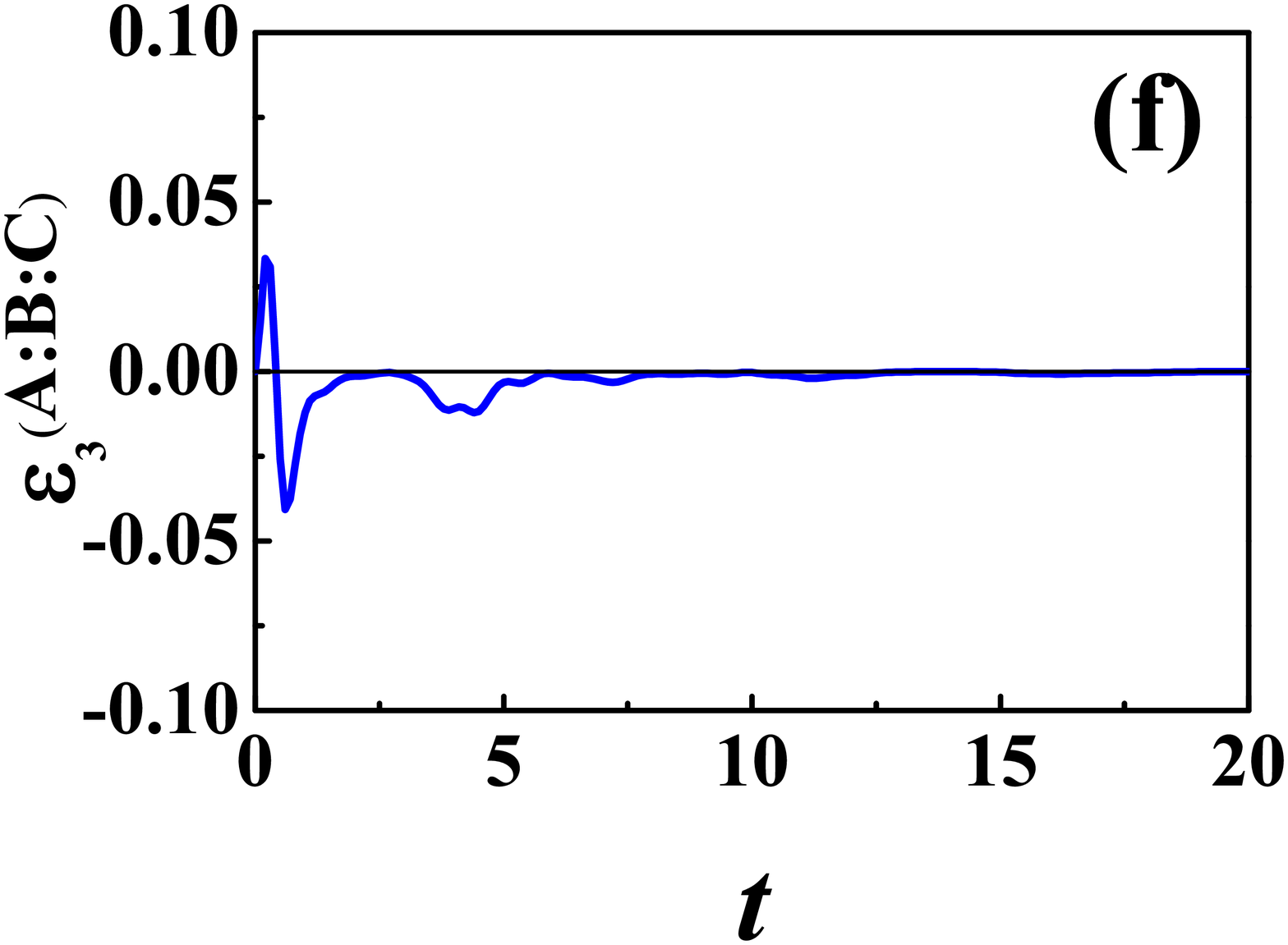}
\caption{TMI and TLN of $XXZ$ chain versus time $t$ in the case of  $L = \sigma ^ z$ for initial state $\left| {00...00} \right\rangle$
and different $\gamma$, (a)-(c) for TMI, (d)-(f) for TLN, and (a), (d) $\gamma  = 1$, (b), (e) $\gamma  = 2$, (c), (f) $\gamma  \to \infty$. The other parameters are $\Gamma  = 0.5$, $N = 7$, $n = 1$.}
\end{figure}

 Then we consider the influences of size of C on TMI and TLN for $XXZ$ chain. In Fig. 11. we plot TMI and TLN as functions of time $t$ for different $n$ and initial state  $\left| {00...00} \right\rangle$ in the case of $L = \sigma ^ z$. It can be seen from Fig. 11 that with the increasing of the spin number of subsystem C, both the maximum absolute values of the negative value for TMI and TLN become larger, and it takes more time for TMI to reach its steady value or for TLN to decay to zero. In a word, with the increase in size of C information scrambling can last for longer time and there is more information scrambled. On the other hand, the absolute value of the steady value for TMI  becomes larger with the increasing of $n$, which indicates that the larger size of subsystem C leads to more residual classical information.  As mentioned above the quantum information scrambling can not occur in the case of $L = \sigma ^ -$ for this initial state.
 For initial N\'{E}EL state, we find that it also takes more time for TMI to reach its steady value or for TLN to decay to zero for both dephasing  and dissipation channels for $XXZ$ chain with the increasing of the spin number $n$ of subsystem C. It implies that with the increasing in size of C information scrambling can last longer time for  $XXZ$  chain and this initial state. In addition, for dephasing baths  the absolute value of the steady value for TMI becomes larger and there is more residual classical information  with the increasing of the spin number $n$ of subsystem C for $XXZ$ chain.

\begin{figure}[htbp]
\centering
\includegraphics[width=4cm]{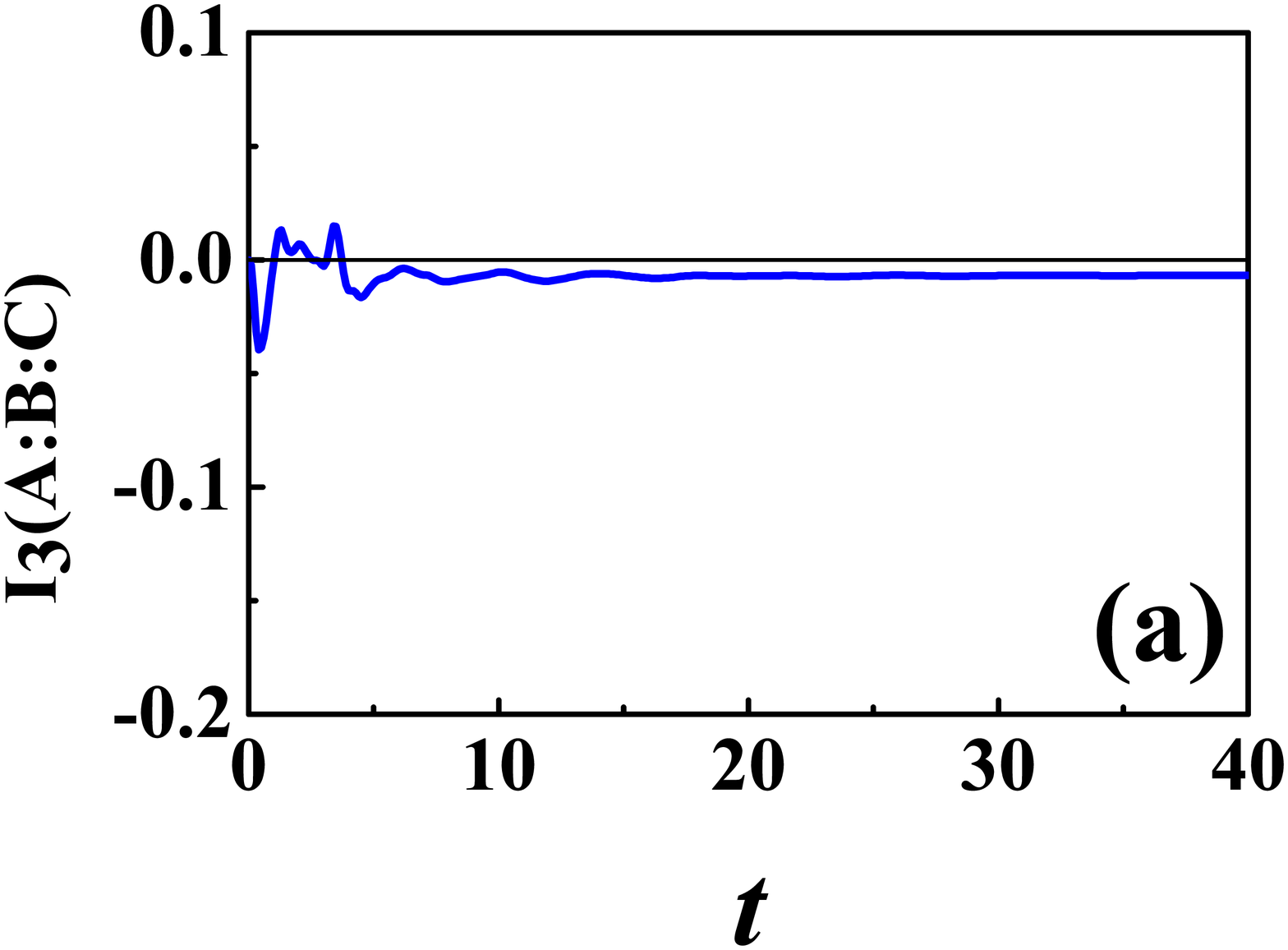}
\includegraphics[width=4cm]{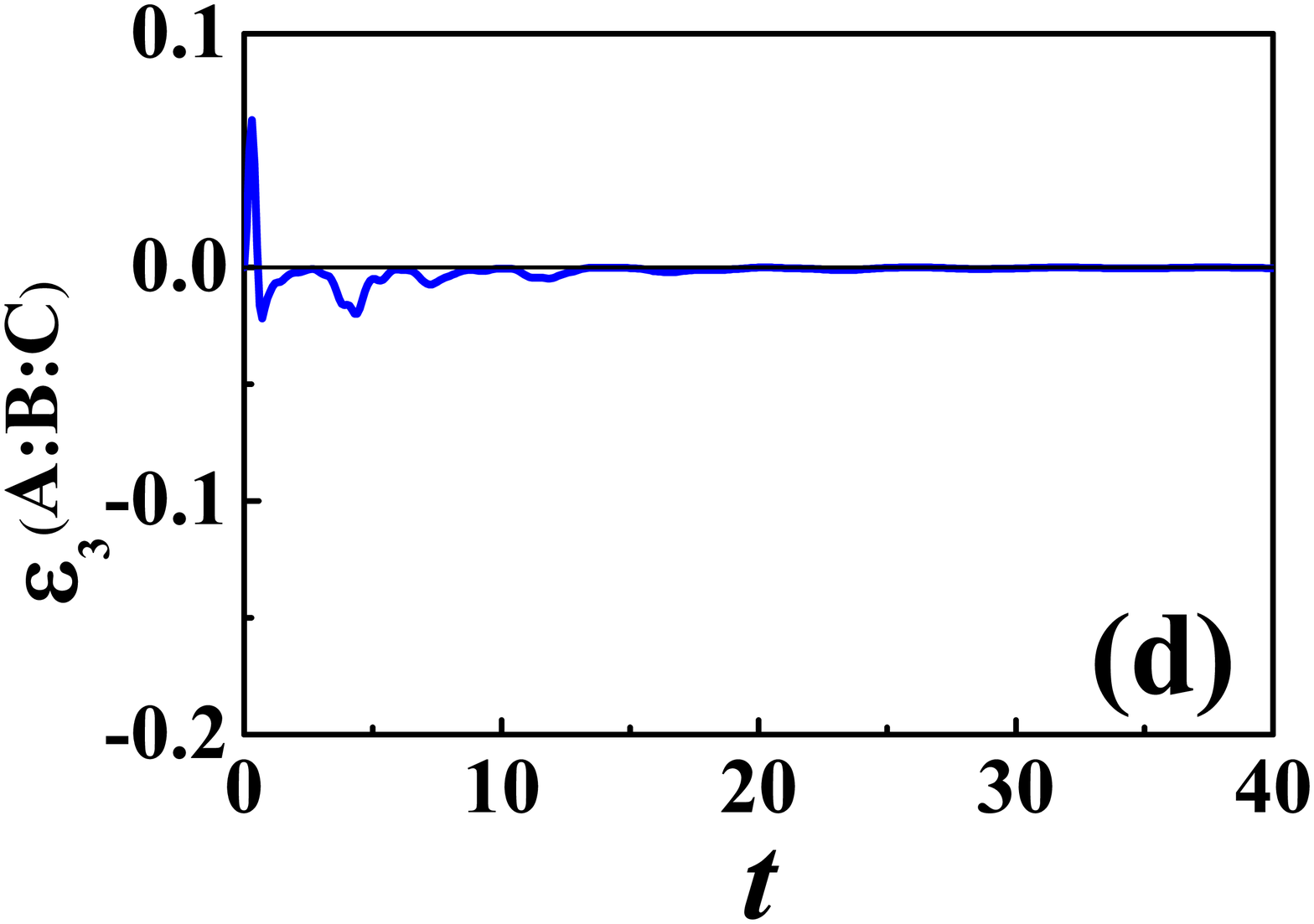}
\includegraphics[width=4cm]{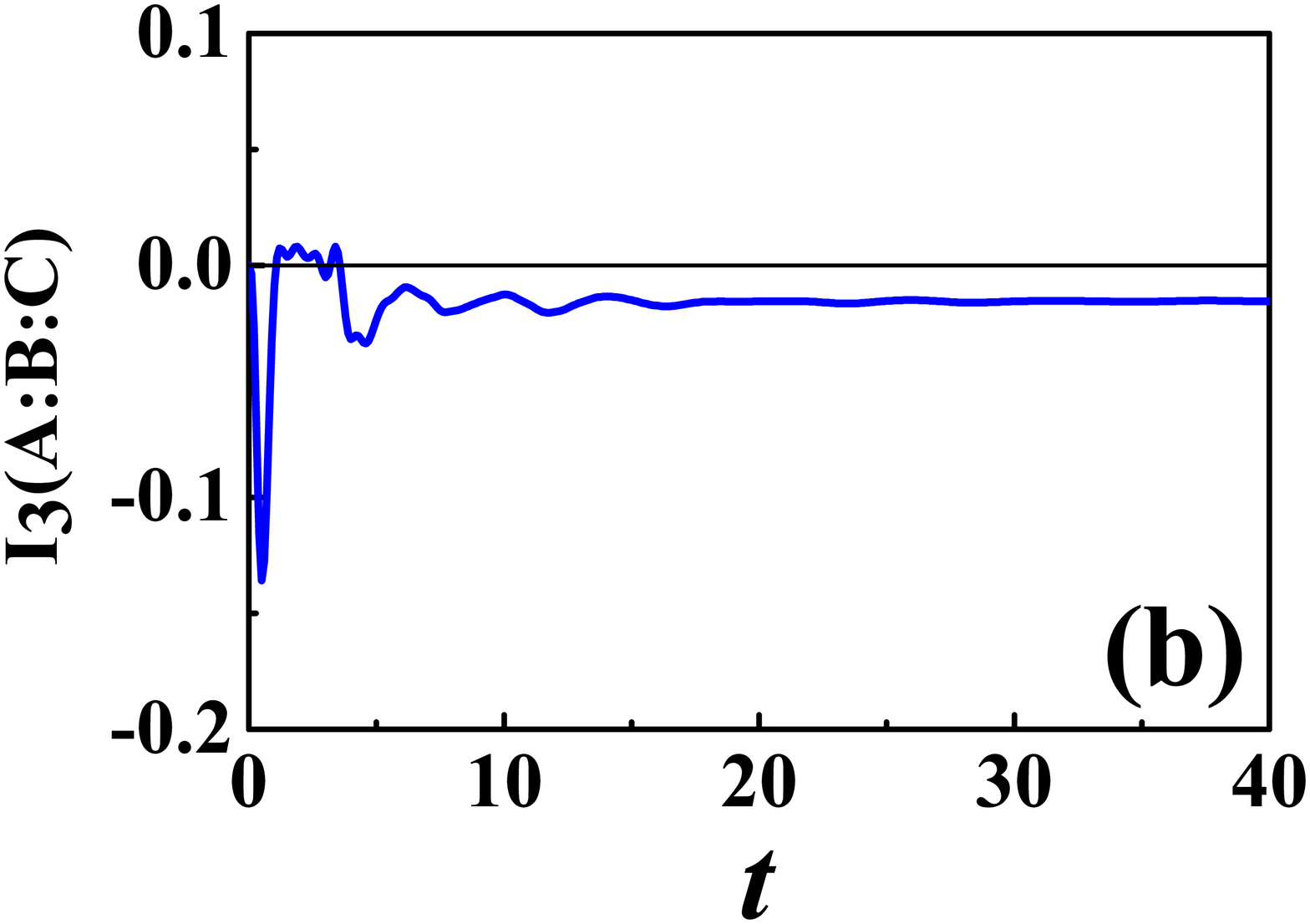}
\includegraphics[width=4cm]{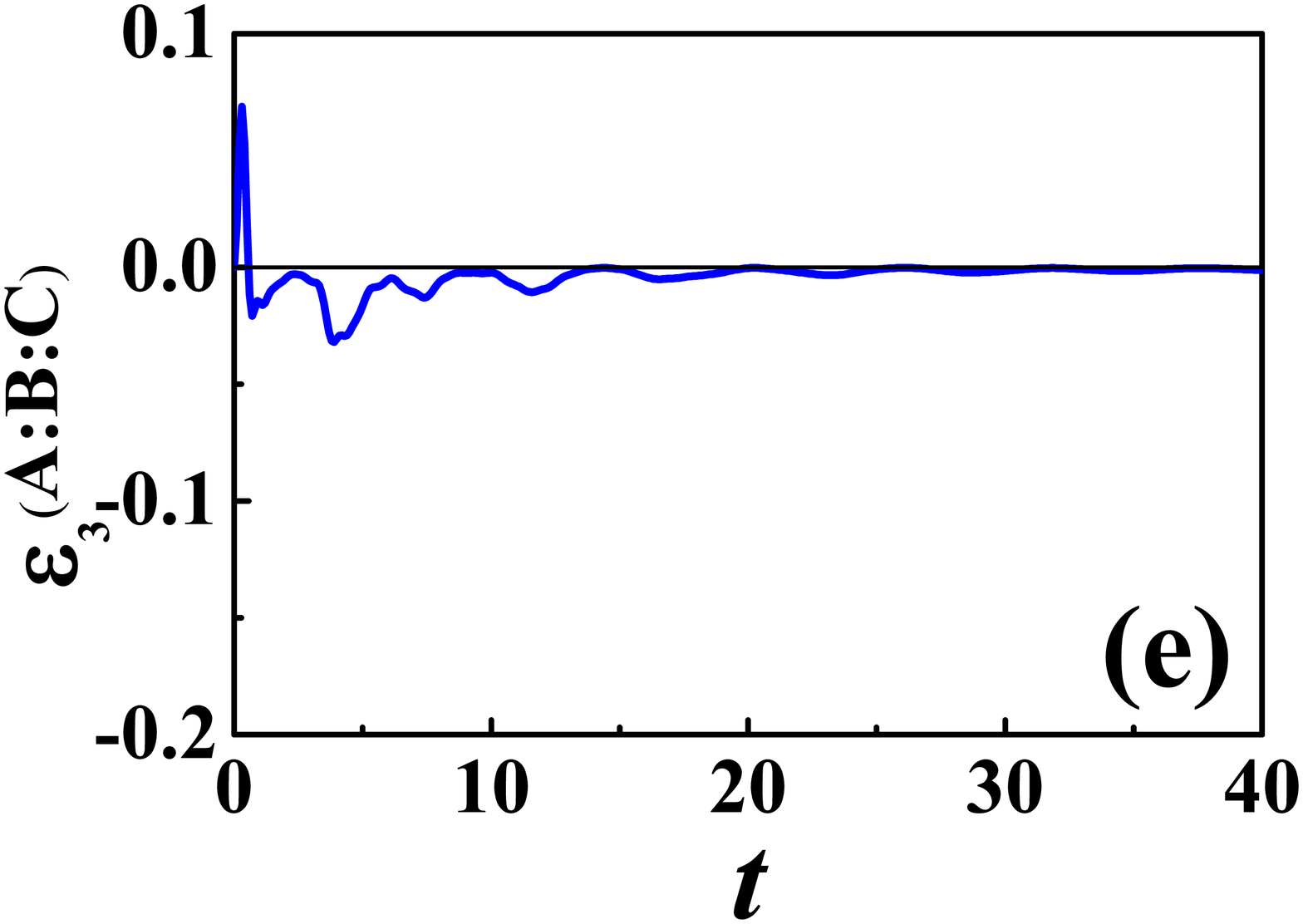}
\includegraphics[width=4cm]{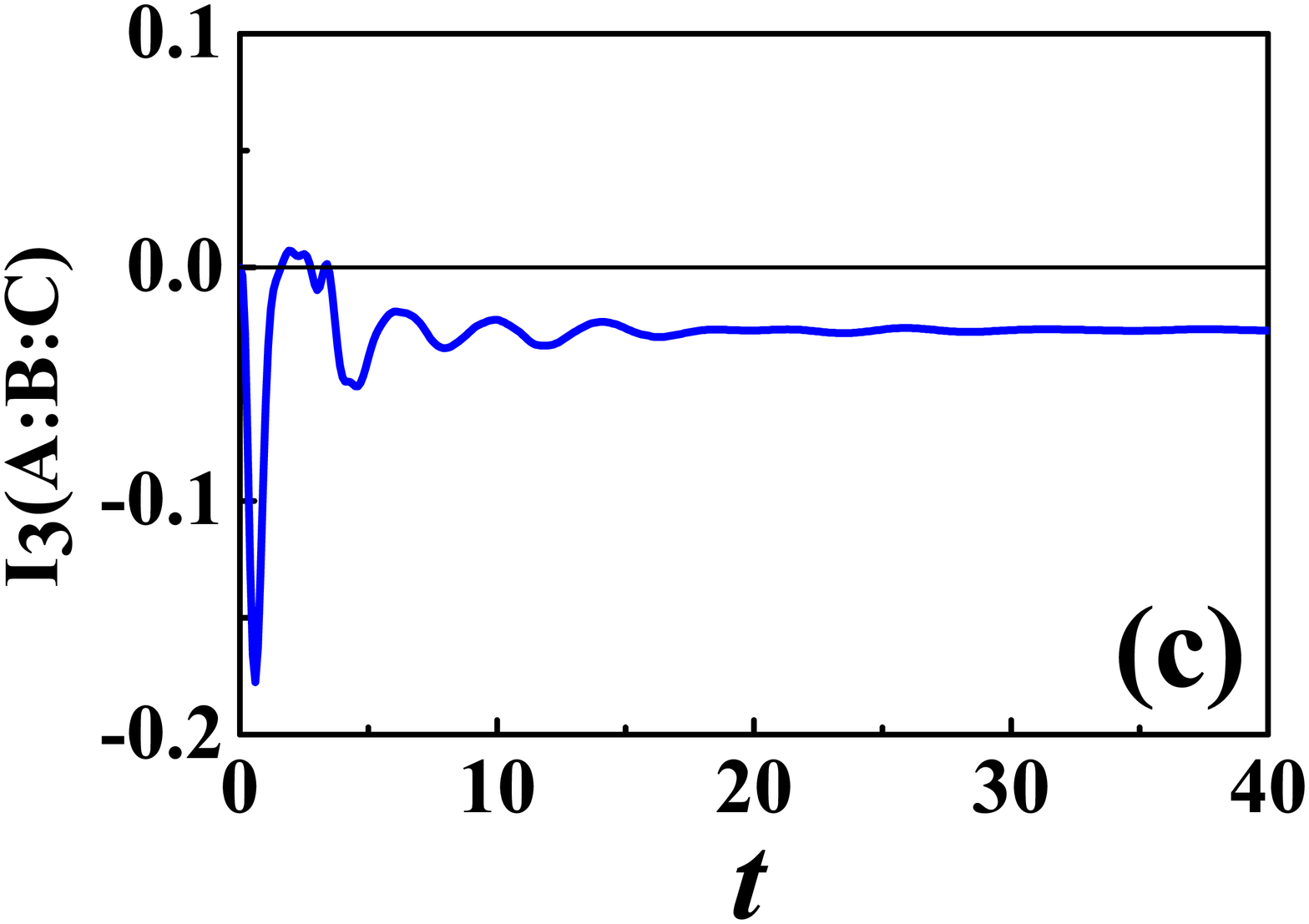}
\includegraphics[width=4cm]{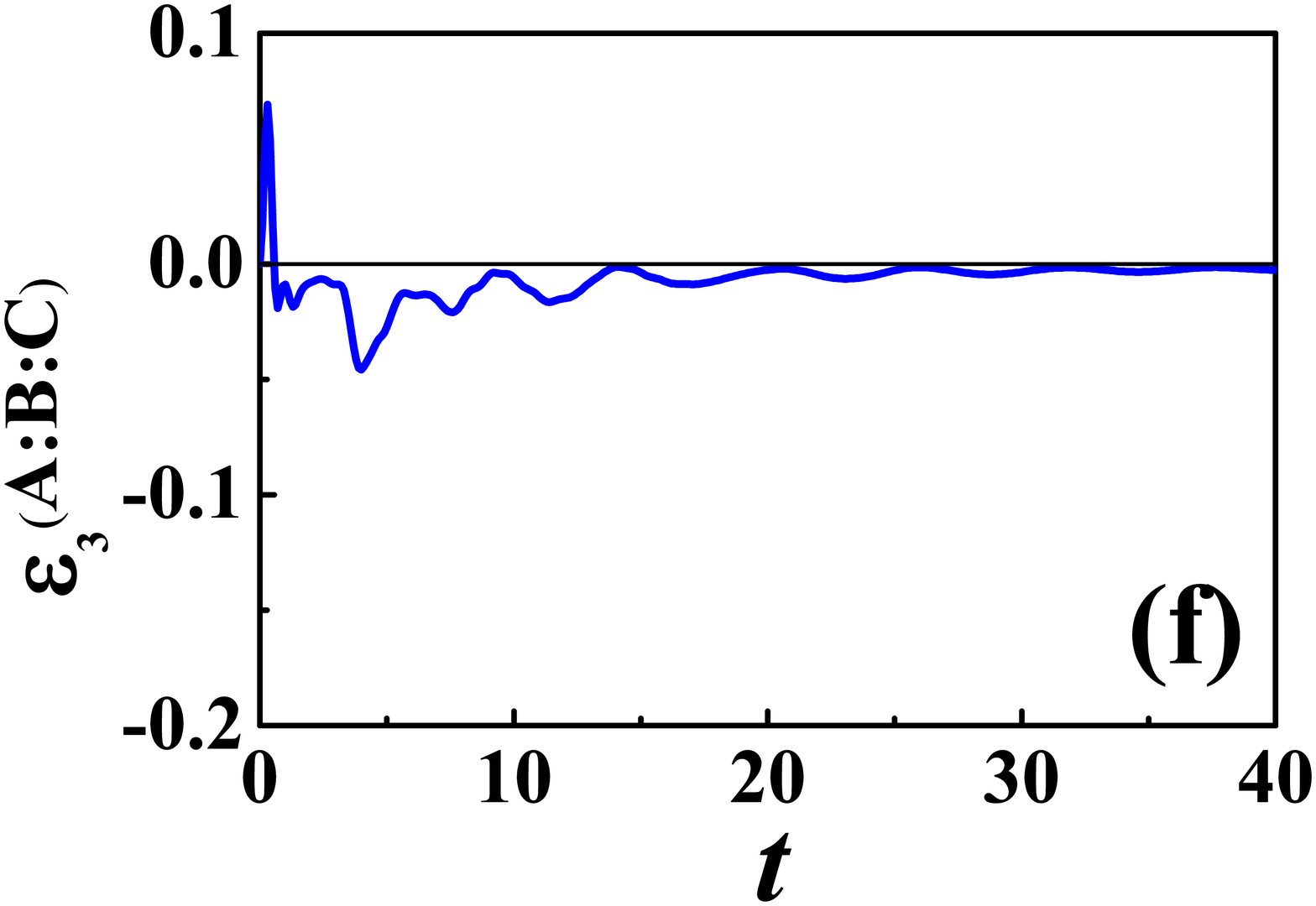}
\caption{TMI and TLN of $XXZ$ chain versus time $t$ in the case of  $L = \sigma ^ z$ for initial state $\left| {00...00} \right\rangle$
and different $n$, (a)-(c) for TMI, (d)-(f) for TLN, and (a), (d) $n  = 1$, (b), (e) $n=2$, (c), (f) $n=3$.  The other parameters are $N = 7$, $\Gamma = 0.5$, $\gamma = 5$.}
\end{figure}

\newpage
\section{THE NUMERICAL RESULTS FOR XX CHAIN}

In this Appendix, we show the numerical results for  $XX$ chain ($\Delta  = 0$), which support the conclusions in the main text. For $XX$ chain, the results for TMI and TLN are almost the same as those for interacting model ($XXZ$ chain), and only slightly different.
From Figs. 12 and 13, we can see that TMI and TLN for initial N\'{E}EL state are both suppressed in the presence of baths, which are qualitatively the same as those for $XXZ$ chain. However for two different types of system-bath interaction, the result is different from that for $XXZ$ chain. From Figs. 13 (b) and (c), we can see that the time interval that TLN stays negative in the case of  $L = \sigma ^ z$ is larger than that in the case of $L = \sigma ^ -$, which is different from the result for $XXZ$ chain shown in Figs. 3 (b) and (c). It indicates that for $XX$ chain, the excitation  decays faster for $L = \sigma ^ -$ than  the coherence decays for $L = \sigma ^ z$. The detrimental effect of dephasing channel is weaker than that of dissipation channel for $XX$ chain for keeping quantum information scrambling. While the maximum absolute value of the negative value of TLN for $L = \sigma ^ -$ is also larger than that for $L = \sigma ^ z$, which is consistent with the results for $XXZ$  chain.
The results for initial state $\left| {00...00} \right\rangle$ are shown in Figs. 14 and 15, and quantum information scrambling can also occur for  $L = \sigma ^ z$, while for this initial state scrambling can not occur for $L = \sigma ^ -$ which are the same as those for $XXZ$ chain shown in Figs. 4 and 5.  For initial N\'{E}EL state, the beneficial effect of the memory effect of $L = \sigma ^ z$ to the emergence of quantum information scrambling can also be seen in Figs. 16 and 17, which are consistent with those shown in Figs. 6 and 7.
For the same initial state, the effect of non-Markovianity of $L = \sigma ^ -$ for $XX$  chain is also helpful for  keeping information scrambling, which is also consistent with the result for $XXZ$ chain. For initial state $\left| {00...00} \right\rangle$, the beneficial effect of the memory  can also be seen for $L = \sigma ^ z$, and for $L = \sigma ^ -$ both information delocalization and quantum information scrambling can not occur whether in the Markovian or non-Markovian regimes, which are qualitatively the same as the results for $XXZ$ chain.
Moreover, we find the influences of size of C on TMI and TLN for $XX$ chain are also qualitatively the same as those for $XXZ$ chain.  With
the increasing in size of C information scrambling can also last longer time for  initial N\'{E}EL state in both two types of system-bath interaction and for initial state $\left| {00...00} \right\rangle$ in the case of $L = \sigma ^ z$.  For $L = \sigma ^ z$ the absolute value of the steady value for TMI also becomes larger and there is more residual classical information with the increasing of the spin number $n$ of subsystem C.  Quantum information scrambling can not occur in the case of $L = \sigma ^ -$ for initial state $\left| {00...00} \right\rangle$.


 \begin{figure}[h!]

\centering
\includegraphics[width=8.3cm]{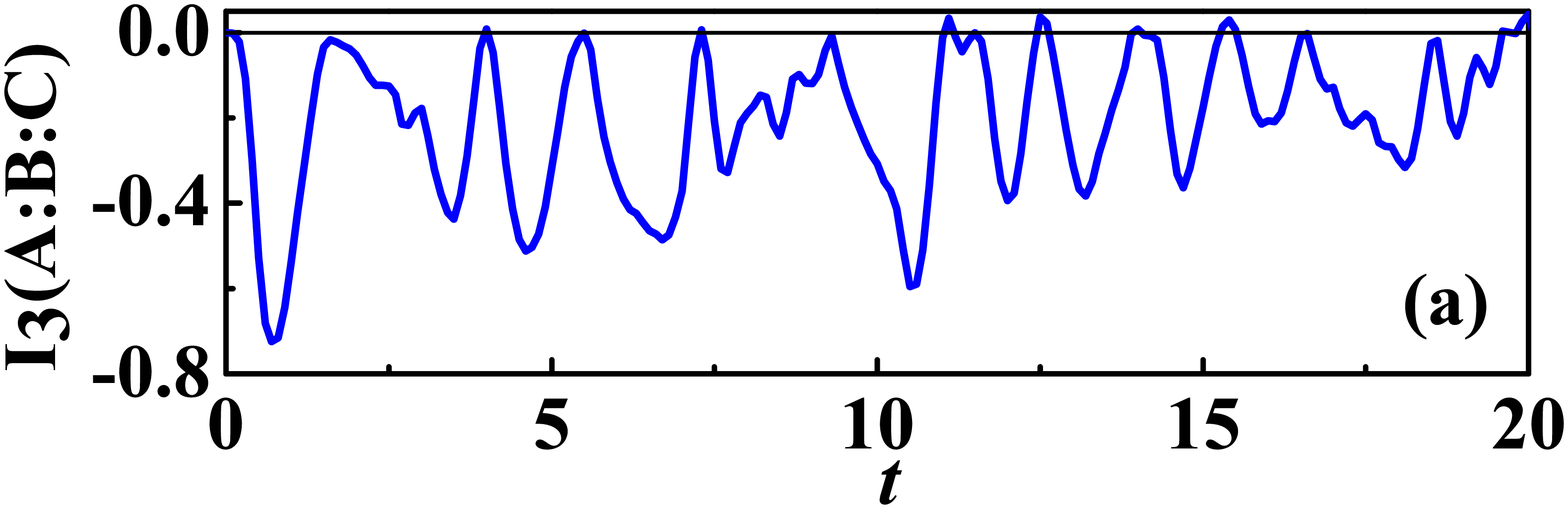}
\centering
\includegraphics[width=8.3cm]{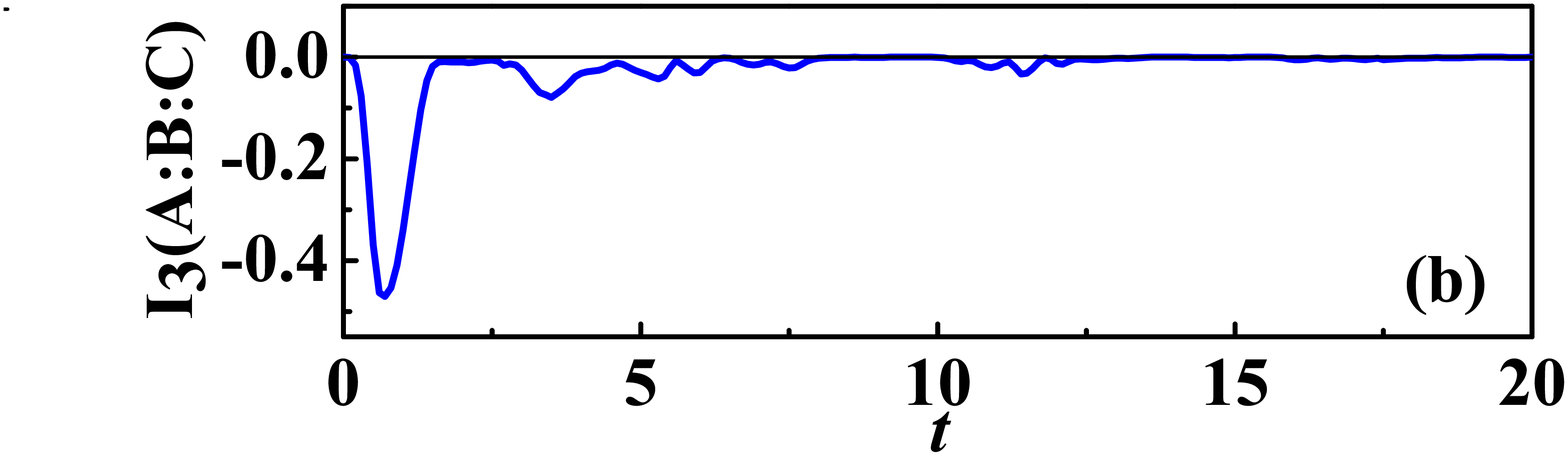}
\centering
\includegraphics[width=8.3cm]{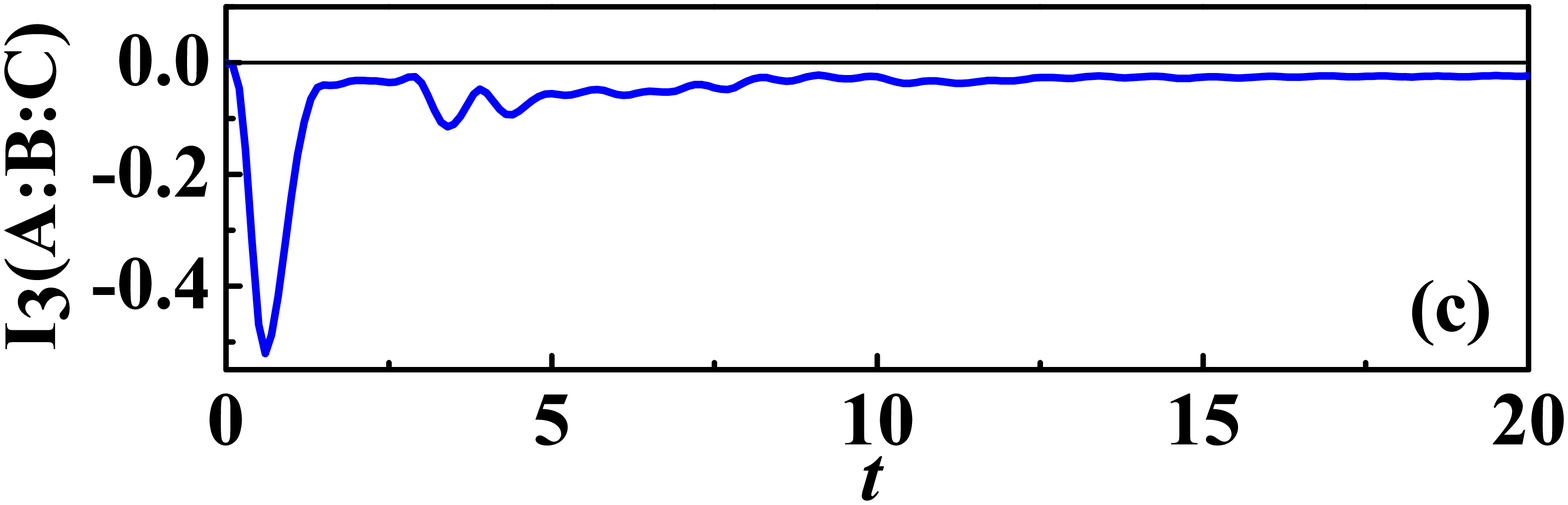}
\caption{TMI of $XX$ chain as a function of time for initial N\'{E}EL state, (a) in the absence of bath ($\Gamma  = 0$), (b) $L = \sigma ^ -$, and (c) $L = \sigma ^ z$. For both (b) and (c), $\Gamma _1  = \Gamma _2  = \Gamma {\rm{ = }}0.5$, and $\gamma _1  = \gamma _2  = \gamma {\rm{ = }}5$. Here $N = 6$, $n = 2$.}
\end{figure}

\begin{figure}[h!]

\centering
\includegraphics[width=8.3cm]{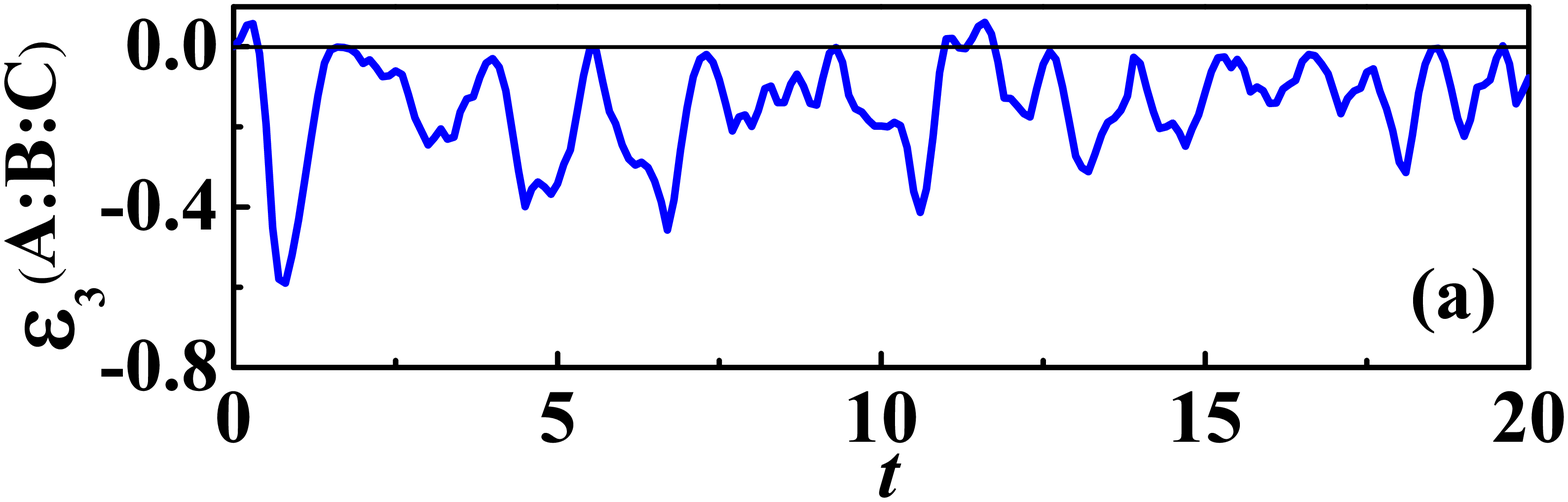}
\centering
\includegraphics[width=8.3cm]{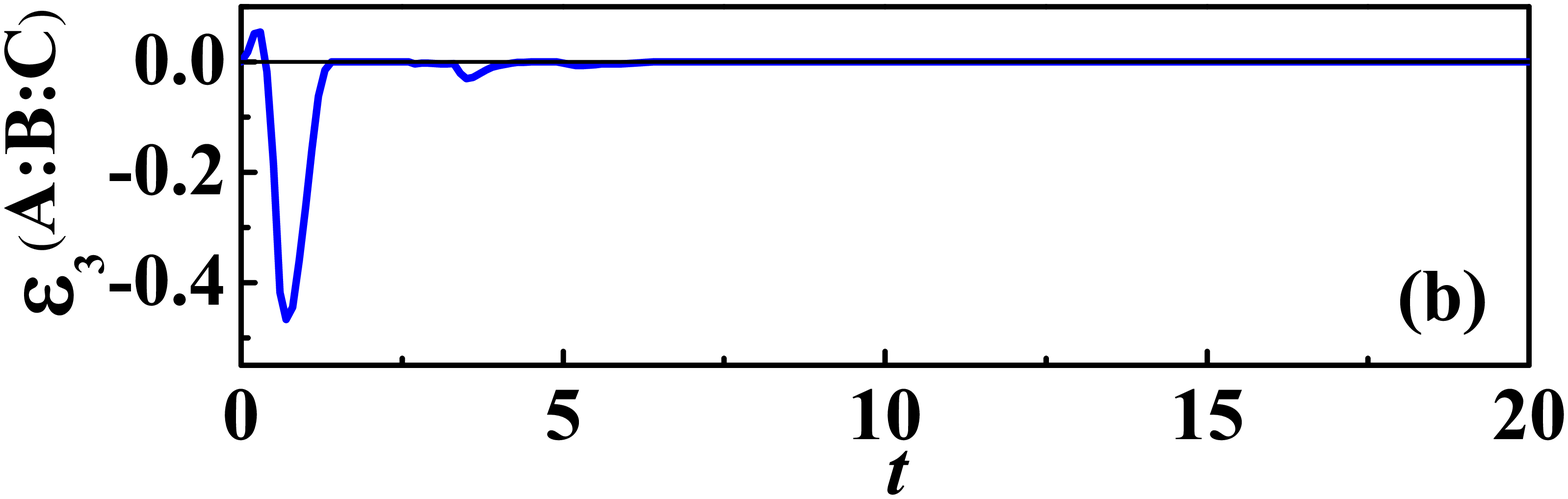}
\centering
\includegraphics[width=8.3cm]{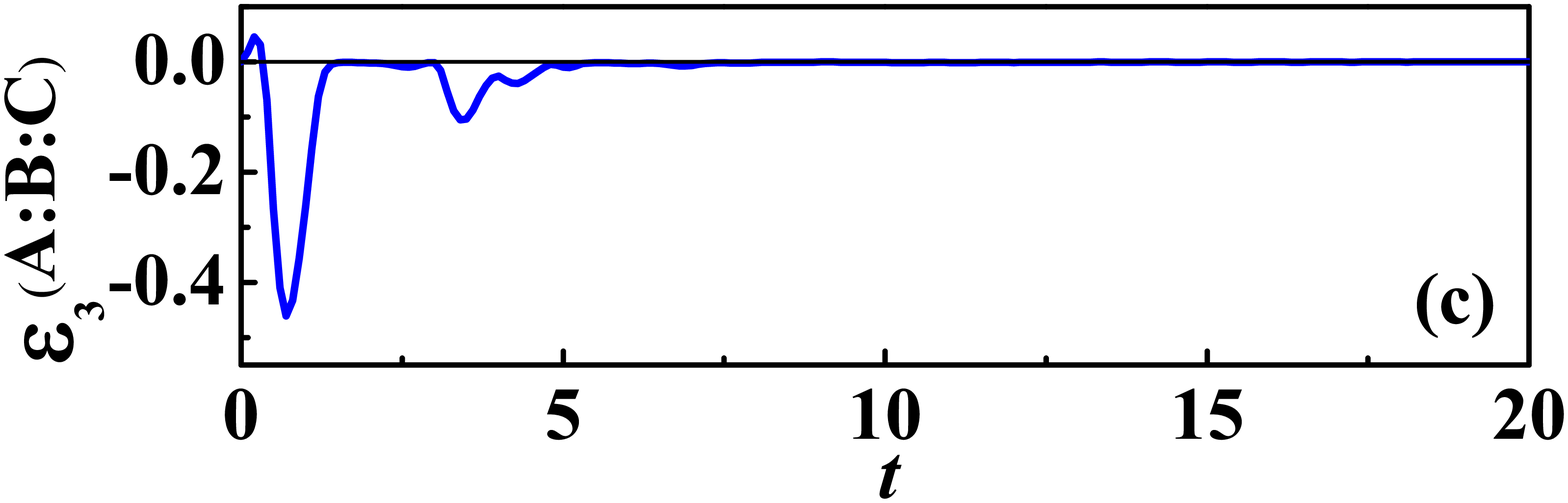}
\caption{TLN of $XX$ chain as a function of time for initial N\'{E}EL state, (a) in the absence of bath ($\Gamma  = 0$), (b) $L = \sigma ^ -$, and (c) $L = \sigma ^ z$. All the parameters are the same as those in Fig. 12.  }
\end{figure}

 \begin{figure}[h!]
\centering
\includegraphics[width=8.3cm]{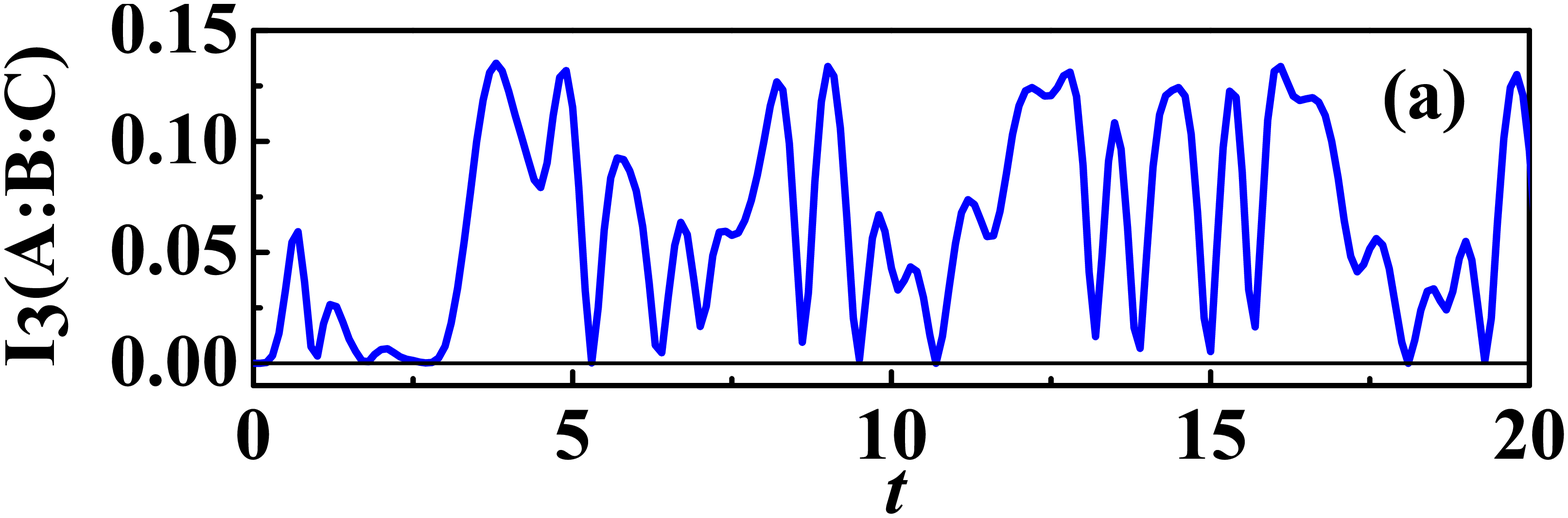}
\includegraphics[width=8.3cm]{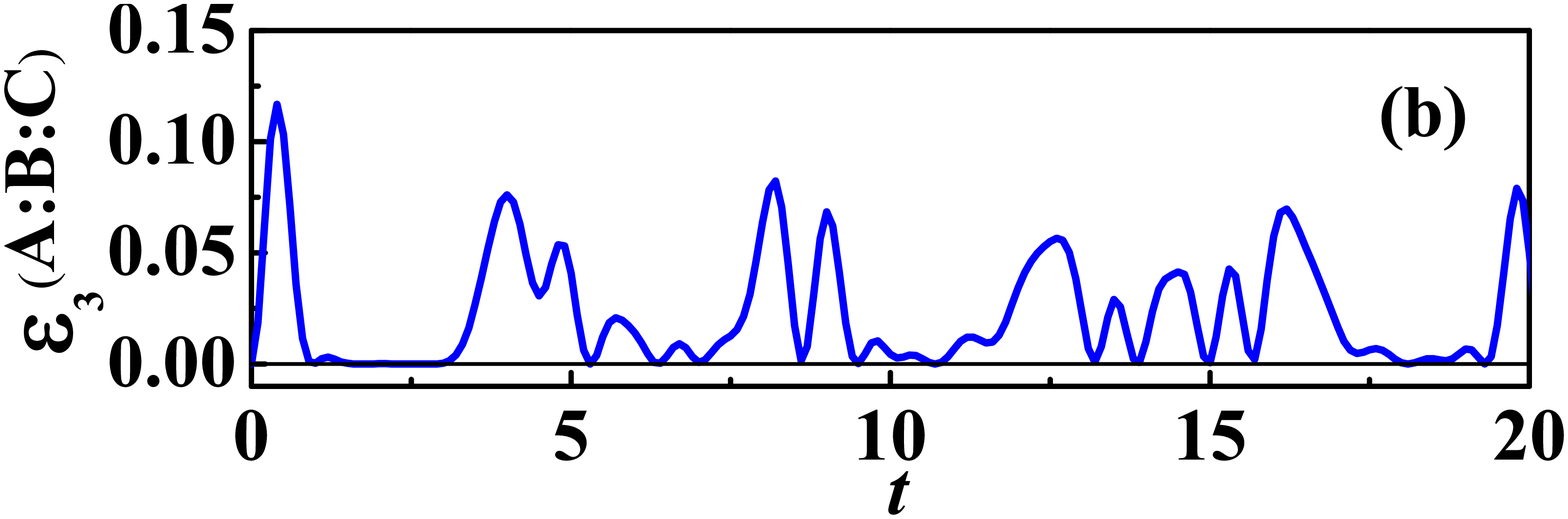}
\caption{TMI and TLN of $XX$ chain as functions of time for initial state $\left| {00...00} \right\rangle$ in the unitary case ($\Gamma  = 0$), (a) for TMI, and (b) for TLN. Here we choose $N = 7$, $n = 1$.}
\end{figure}

\begin{figure}[h!]
\centering
\includegraphics[width=8.3cm]{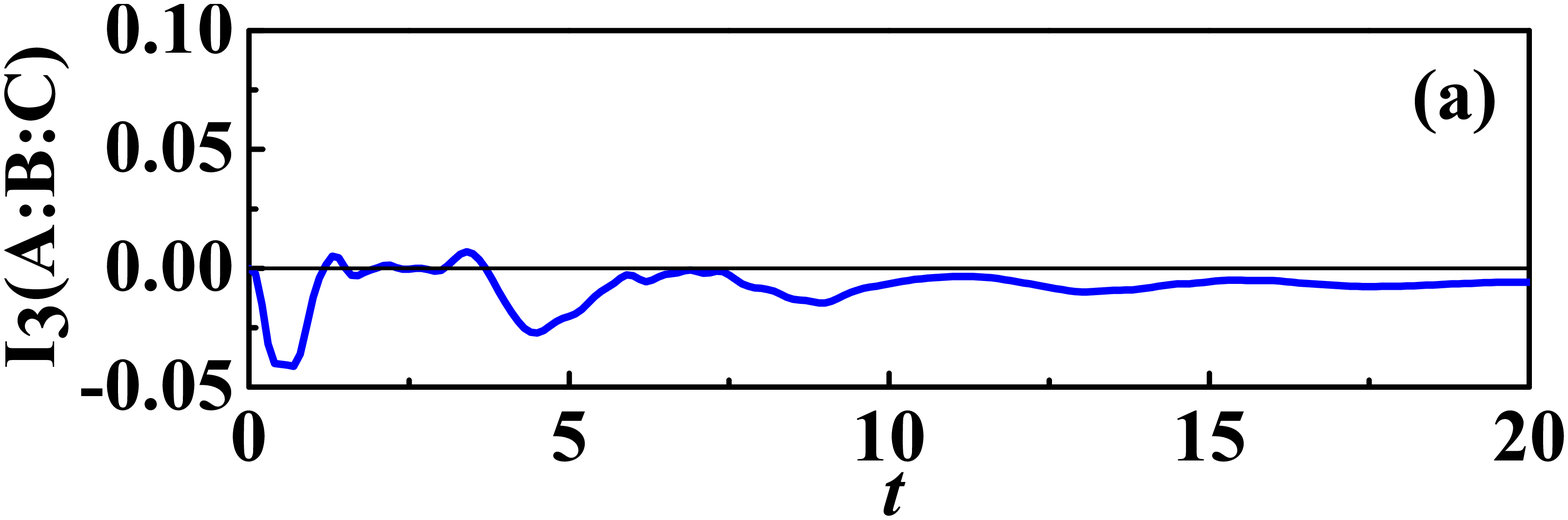}
\includegraphics[width=8.3cm]{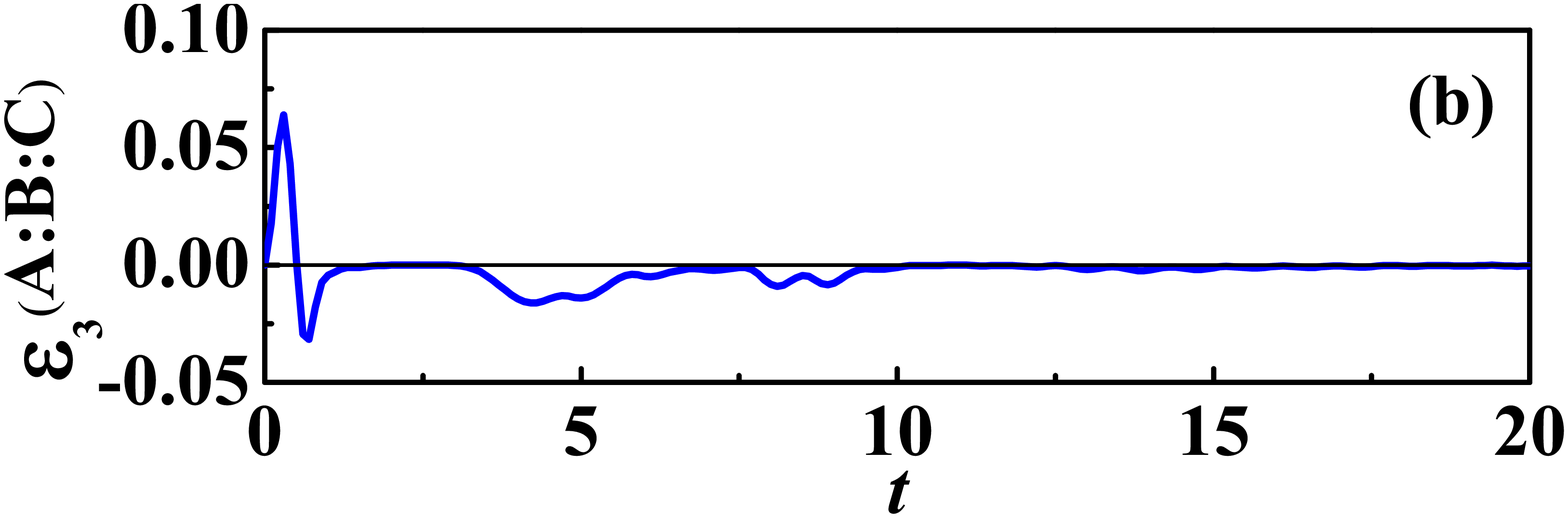}
\includegraphics[width=8.3cm]{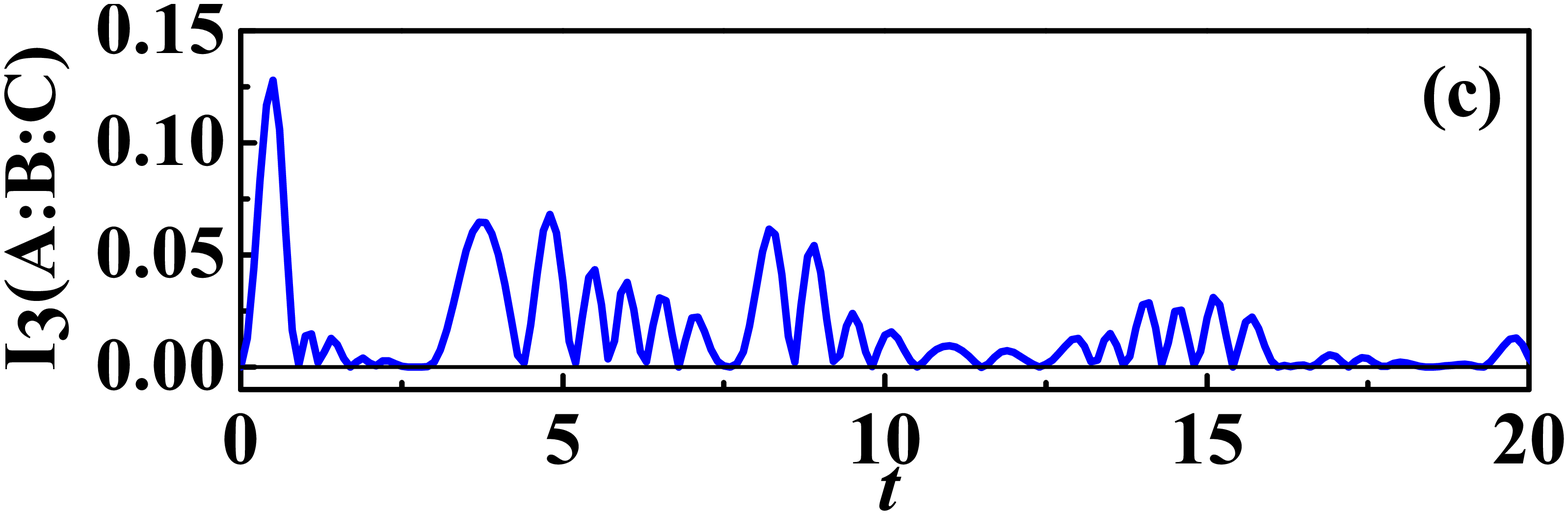}
\includegraphics[width=8.3cm]{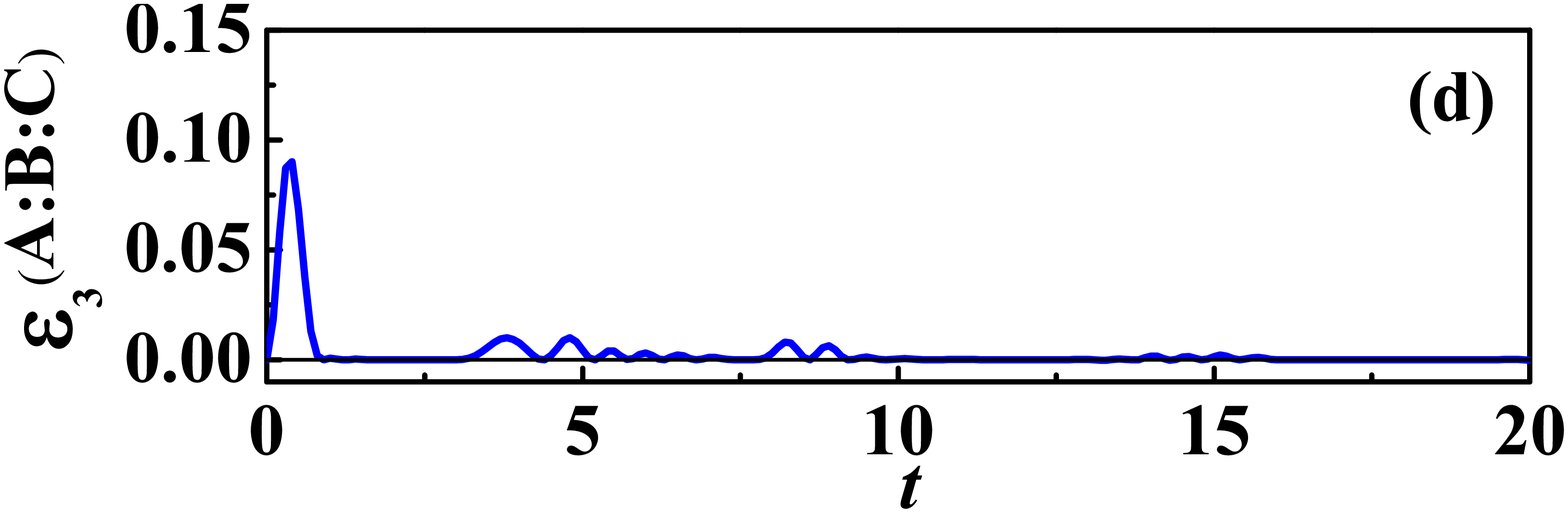}
\caption{TMI and TLN of $XX$ chain as functions of time for initial state $\left| {00...00} \right\rangle$, (a) and (b) for TMI and TLN in the case of  $L = \sigma ^z$ respectively; (c) and (d) for TMI and TLN in the case of  $L = \sigma ^-$ respectively. Here $N = 7$, $n = 1$ and  the other parameters are the same as those in Fig. 12.}
\end{figure}
\begin{figure}[h!]
\centering
\includegraphics[width=8.3cm]{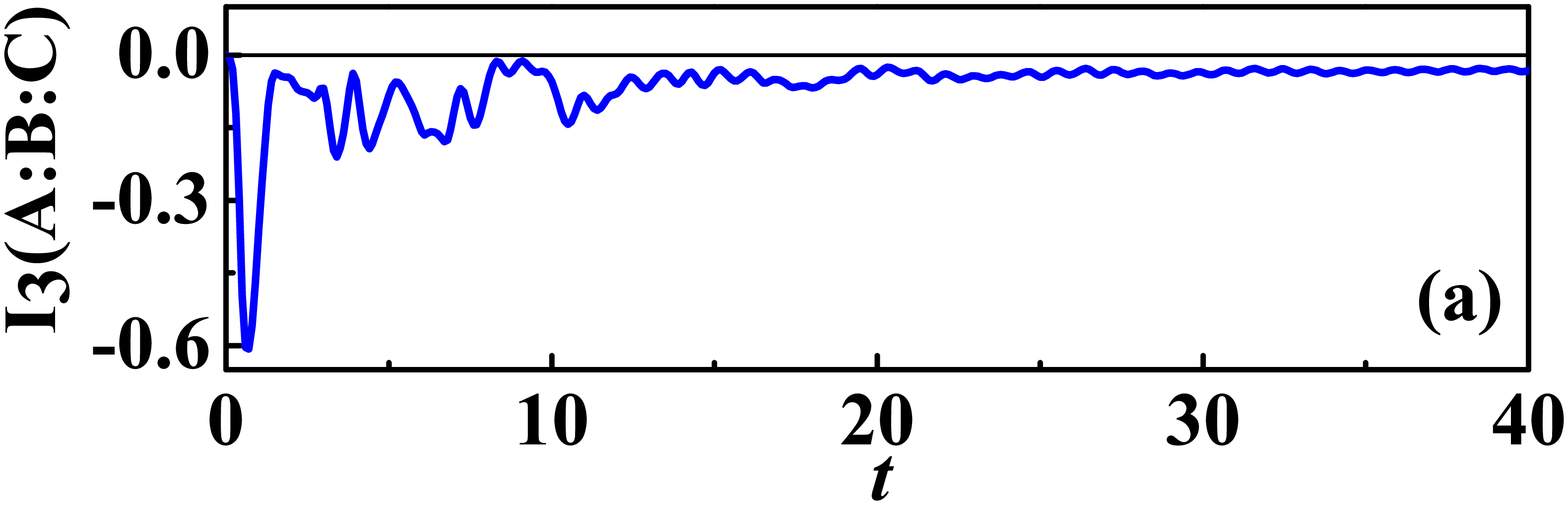}
\includegraphics[width=8.3cm]{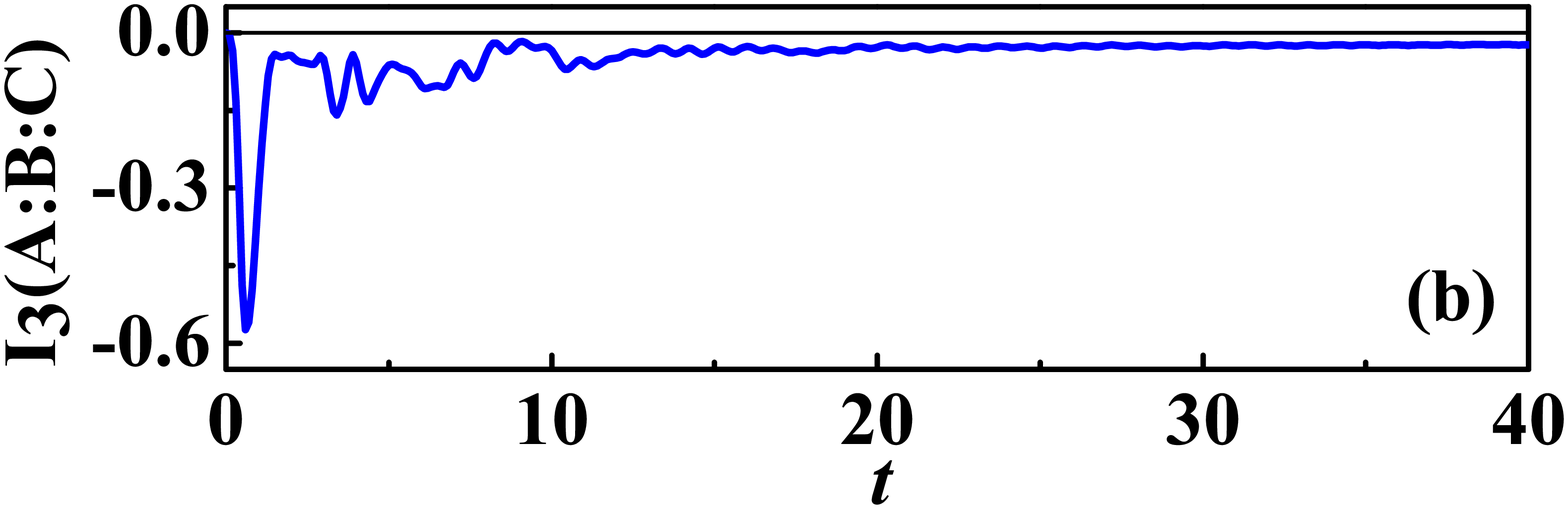}
\includegraphics[width=8.3cm]{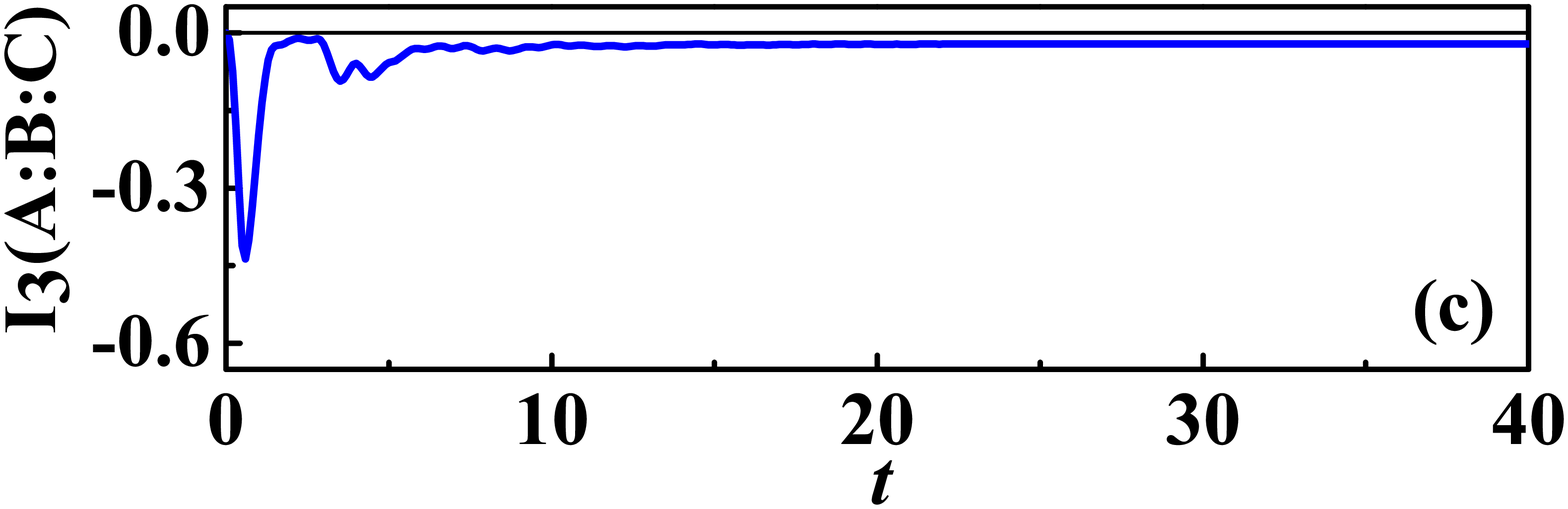}
\caption{TMI of $XX$ chain versus time $t$ in the case of  $L = \sigma ^ z$ for initial N\'{E}EL state and different $\gamma$, (a) $\gamma  = 1$, (b) $\gamma  = 2$, and (c) $\gamma  \to \infty$. The other parameters are $\Gamma  = 0.5$, $N = 6$, $n = 2$.}
\end{figure}

\begin{figure}[h!]
\centering
\includegraphics[width=8.3cm]{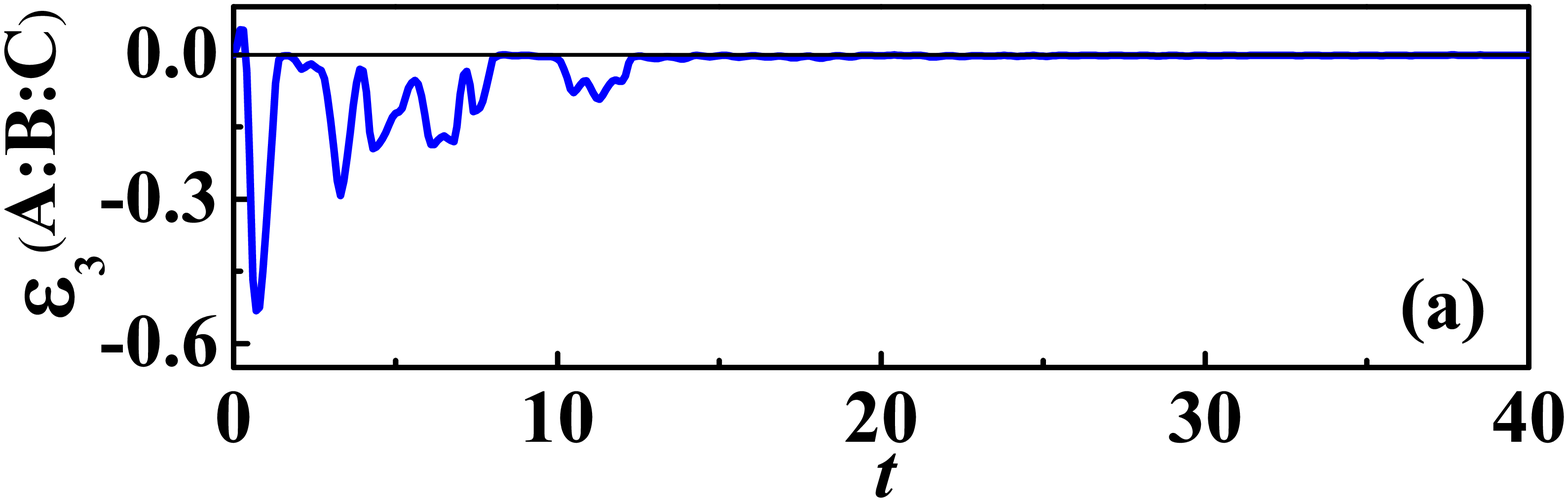}
\includegraphics[width=8.3cm]{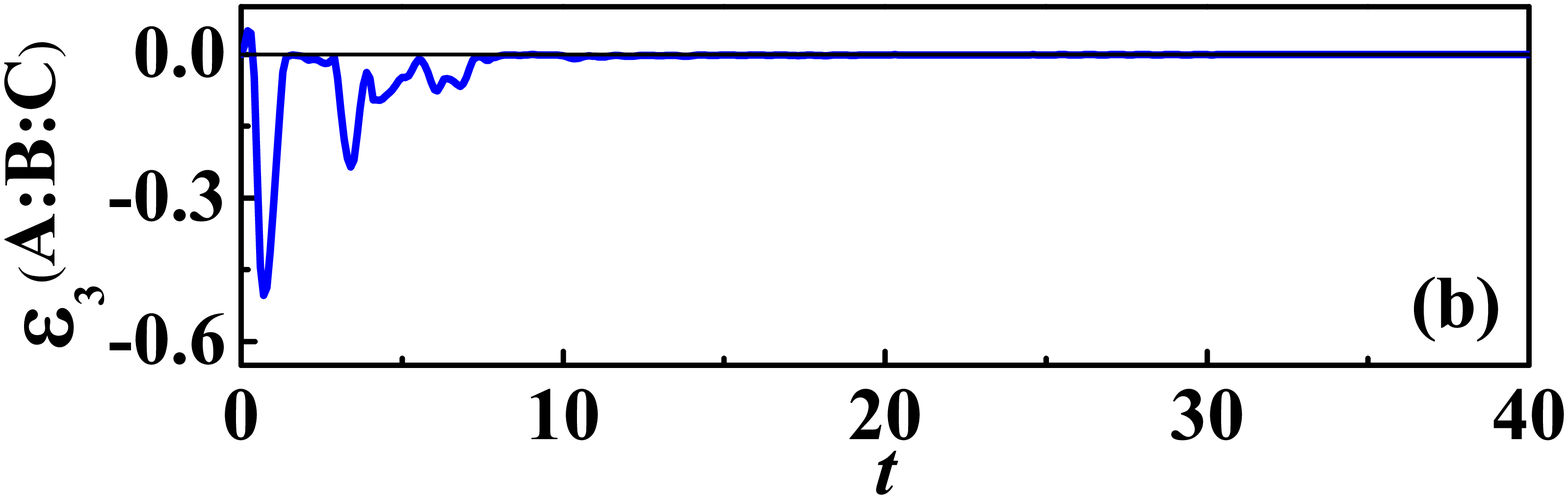}
\includegraphics[width=8.3cm]{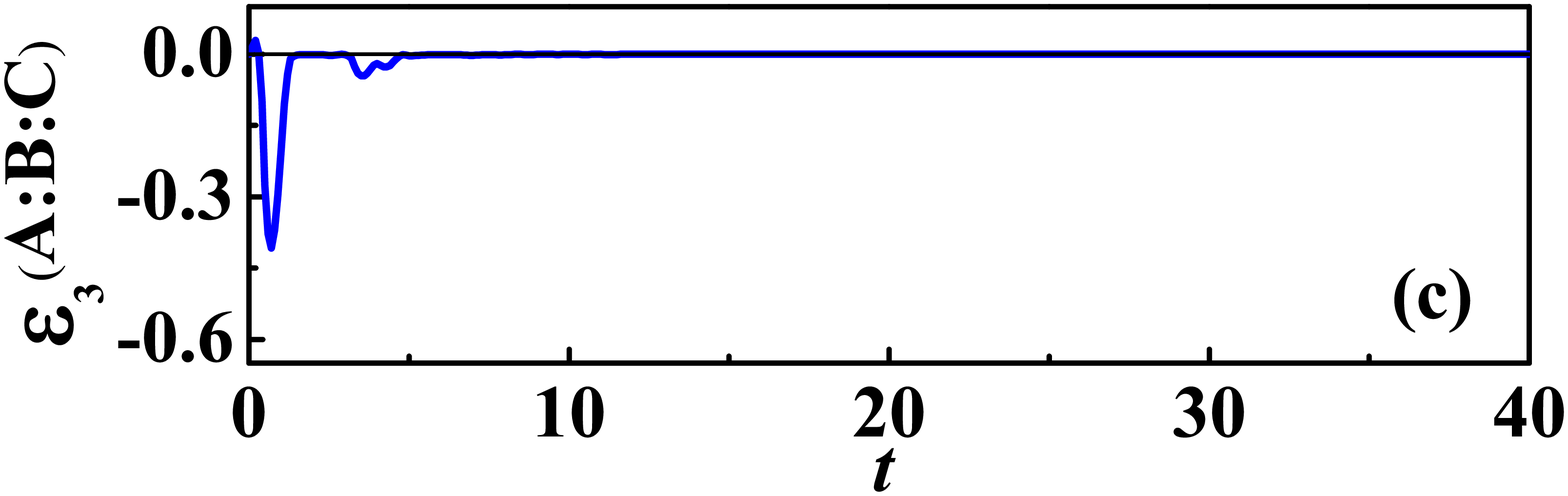}
\caption{TLN of $XX$ chain versus time $t$ in the case of  $L = \sigma ^ z$ for initial N\'{E}EL state and different $\gamma$, (a) $\gamma  = 1$, (b) $\gamma  = 2$, and (c) $\gamma  \to \infty$. The other parameters are the same as those in Fig. 16.}
\end{figure}

\clearpage
\nocite{*}
\bibliographystyle{apsrev4-1}

\end{document}